\documentclass[aip,preprint]{revtex4-1}

\usepackage{mathrsfs}
\usepackage{amsmath}
\usepackage{stmaryrd}
\usepackage{bbding}
\usepackage{dcolumn}
\usepackage{graphicx}
\usepackage{amsfonts}
\usepackage{amssymb}
\usepackage{psfrag}
\usepackage{wrapfig}
\usepackage{subfigure}
\usepackage{makeidx}
\usepackage{bm}
\usepackage{epsf}
\usepackage{epsfig}
\usepackage{setspace}
\usepackage{graphicx}
\usepackage{epstopdf}
\usepackage{psfrag}
\usepackage{subfigure}
\usepackage{multirow}
\usepackage{diagbox}
\usepackage{verbatim}
\usepackage{makecell}
\usepackage{color}
\usepackage{lineno,hyperref}

\begin{document}
	
	
	\title{Physical Modeling and Numerical Studies of Three-dimensional Non-equilibrium Multi-temperature Flows}
	
	\author{Guiyu Cao}
	\email{gcaoaa@connect.ust.hk}
	\affiliation{Department of Mathematics, Hong Kong University of Science and Technology, Clear Water Bay, Kowloon, Hong Kong}
	\author{Hualin Liu}
	\email{hualinliu@zju.eud.cn}
	\affiliation{College of Aeronautics and Astronautics, Zhejiang University, Hangzhou, Zhejiang 310058, China}
	\author{Kun Xu}
	\email{makxu@ust.hk}
	\affiliation{Department of Mathematics, Hong Kong University of Science and Technology, Clear Water Bay, Kowloon, Hong Kong}
	\affiliation{Department of Mechanical and Aerospace Engineering, Hong Kong University of Science and Technology, Clear Water Bay, Kowloon, Hong Kong}
	
	\date{\today}
	
	\begin{abstract}
		For increasingly rarefied flowfields, the Navier-Stokes (NS) equations lose accuracy partially due to the single temperature approximation. To overcome this barrier, a continuum multi-temperature model based on the Bhatnagar-Gross-Krook (BGK) equation coupled with the Landau-Teller-Jeans relaxation model has been proposed for two-dimensional hypersonic non-equilibrium multi-temperature flow computation. In recent study, a two-stage fourth-order gas-kinetic scheme (GKS) has been developed for equilibrium flows, which achieves a fourth-order accuracy in space and time as well as high efficiency and robustness. In this paper, targeting for accurate and efficient simulation of  multi-temperature non-equilibrium flows, a high-order three-dimensional multi-temperature GKS is implemented under the two-stage fourth-order framework, with the fourth-order Simpson interpolation rule for the newly emerged source term. Simulations on decaying homogeneous isotropic turbulence, low-density nozzle flow, rarefied hypersonic flow over a flat plate, and type IV shock-shock interaction are used to validate the multi-temperature model through the comparison with experimental measurements. 
		The unified gas kinetic scheme (UGKS) results, and the Direct simulation Monte Carlo (DSMC) solutions will be used as well in some cases for validation. Computational results not only confirm the high-order accuracy and quite robustness of this scheme, but also show the significant improvement on computational efficiency compared with UGKS and DSMC, especially in the near continuum flow regime.
	\end{abstract}
	
	
	\maketitle

\section{Introduction}
The classification of flow regimes is based on the Knudsen number $Kn$, which is defined as the ratio of the molecular mean free path over a characteristic length scale of the system. The whole flow regime is roughly divided into continuum flow regime ($Kn \le 0.01$), continuum-transition regime ($0.01 < Kn \le 10$), and free molecular regime ($Kn > 10$). The Navier-Stokes (NS) equations with linear relations between stress and strain and the Fourier's laws are adequate to model the equilibrium flow in the continuum flow regime. For non-equilibrium flow in the continuum-transition regime, the Navier-Stokes equations are well known to be inadequate. However, this continuum-transition regime is important for many scientific and practical engineering applications, such as the simulation of micro-scale flows and space exploration vehicles\cite{ivanov1998computational}. Therefore, accurate models with reliable solutions and lower computational costs for non-equilibrium flow are useful for solving the non-equilibrium flow problem in the near continuum regime.

Available numerical schemes for simulating non-equilibrium flow can be classified into particle method and deterministic method. Direct simulation Monte Carlo (DSMC) \cite{bird1978monte,bird1994molecular} uses probabilistic simulation to solve the Boltzmann equation, which is a representative of particle method and is widely used for rarefied low simulations. However, in the continuum-transition regime, DSMC requires a great amount of particles and  the cell size and time step are limited by the particle mean free path and mean collision time, which is very expensive both in the memory cost and computational time. The deterministic method, such as Discrete Velocity Methods (DVM) or Discrete Ordinate Method (DOM) \cite{yang1995rarefied, mieussens2000discrete, kolobov2007unified, li2009gas}, solve the Boltzmann or model equations directly with the discretization of particle velocity space. In the continuum-transition regime, the cell size and time step are also constrained by the particle mean free path and mean collision time, make these methods be prohibitively expensive. Recently, the multi-scale numerical scheme unified gas kinetic scheme (UGKS) \cite{xu2010unified,huang2013unified,liu2014unified,xu2015direct} has been developed successfully for monatomic and diatomic gases for entire Knudsen number flow. Different from the splitting process used in DSMC and DVM/DOM methods, the distinguishable feature of UGKS is the coupling of the particle transport and collision, which makes the grid size and time step used in UGKS are not limited by the particle mean free path and collision time, such as those imposed in DSMC and DVM methods. Even though UGKS is the most efficient method for whole flow regime simulation currently, in view of a considerable number of discrete velocity points to be updated, it is still expensive in the near continuum flow regime than those based on the macroscopic equations. At the same time, for smooth flow, shch as those in the boundary layer, a high-order scheme is preferred to get accurate solutions. However, most schemes for the rarefied flow, such as DSMC, DVM/DOM, and UGKS methods, have only at most second-order accuracy.

To study non-equilibrium flow efficiently, an extended Bhatnagar-Gross-Krook (BGK) model coupled with the Landau-Teller-Jeans relaxation model has been proposed for two-dimensional non-equilibrium multi-temperature flow computation \cite{xu2006continuum, xu2008multiple}. In the continuum flow regime, the corresponding kinetic scheme goes back automatically to the BGK-NS method. On the other hand, this kinetic scheme solves the non-equilibrium translational and rotational flow quite efficiently in the near continuum regime. In recent study, an accurate and robust two-stage fourth-order gas-kinetic scheme (GKS) \cite{li2016two, pan2016efficient} has been developed for equilibrium flows, which achieves a fourth-order accuracy in space and time, and shows high efficiency and robustness from smooth flow to shock problem. In view of smooth equilibrium region appearing in the  non-equilibrium multi-temperature flows, a high-order non-equilibrium GKS based on extended BGK method is preferred for simulating multi-temperature flow efficiently and accurately. In current study, this high-order non-equilibrium GKS is implemented under the previous two-stage fourth-order framework for three-dimensional multi-temperature flows, and the source term is dealt with fourth-order Simpson interpolation rule. Numerical tests from smooth decaying homogeneous isotropic turbulence to challenging hypersonic type IV shock-shock interaction validate current high-order non-equilibrium GKS. This high-order non-equilibrium GKS not only preserves high accuracy and quite robustness through numerical cases, but also shows the significant improvement on computational efficiency  in near continuum flow region.

In this paper, details on current extended kinetic model and corresponding macroscopic equations are presented in Section 2. Section 3 gives  the construction of this high-order non-equilibrium numerical scheme under two-stage fourth-order framework for solving this extended kinetic model. This is followed by the results and discussion of the non-equilibrium multi-temperature flow computations in Section 4. Discussion and conclusion are shown in the final section.

\newpage

\section{Gas-kinetic models and macroscopic governing equations for diatomic gas}
In this section, the extended kinetic model and its derived macroscopic equations in three dimension for diatomic gases are presented.

\subsection{Equilibrium translational and rotational temperature model}
By modeling the time evolution of a gas distribution function resulting from the free transport and binary elastic collision, the Boltzmann equation has been constructed for monotonic dilute gas. The simplification of the Boltzmann equation given by the BGK model has the following form \cite{bhatnagar1954model},
\begin{align}
    \frac{\partial f}{\partial t} + u \frac{\partial f}{\partial x} + v \frac{\partial f}{\partial y}  + w \frac{\partial f}{\partial z}=  \frac{g - f}{\tau},
    \label{boltzmann_bgk_eq}
\end{align}
where $f$ is the number density of molecules at position $(x,y,z)$ and particle velocity $(u,v,w)$ at time $t$. The left side of the Eq.(\ref{boltzmann_bgk_eq}) denotes the free transport, and the right hand side represents the collision term. The relation between distribution function $f$ and macroscopic variables, such as mass, momentum, energy and stress, can be obtained by taking moments of the distribution function. The collision operator in BGK model shows simple relaxation process from $f$ to a local equilibrium state $g$, with a characteristic time scale $\tau$ related to the viscosity and heat conduction coefficients. The local equilibrium state is a Maxwellian distribution,
\begin{align}
    g = \rho (\frac{\lambda}{\pi})^{\frac{K + 3}{2}}  e^{- \lambda [(u - U)^2 + (v - V)^2 + (w - W)^2 + \xi^2]},
    \label{maxwellian}
\end{align}
where $\rho$ is the density, $(U,V,W)$ are the macroscopic fluid velocity in the $x-$,$y-$ and $z-$ directions. Here $\lambda = m/2kT$, $m$ is the molecular mass, $k$ is the Boltzmann constant, and $T$ is the temperature. For three-dimensional equilibrium diatomic gas, the total number of degrees of freedom $K = 2$, the internal variable $\xi$ accounts for the rotational modes as $\xi^2 = \xi_1^2 + \xi_2^2$, and the specific heat ratio $\gamma = (K + 5)/(K + 3)$ is determined.

\newpage

Based on the above BGK model as Eq.(\ref{boltzmann_bgk_eq}), the Euler equations can be obtained for a local equilibrium state with $f = g$. On the other hand, for the Navier-Stokes equations, the stress and Fourier heat conduction terms can be derived with the Chapman-Enskog expansion \cite{chapman1990mathematical} truncated to the $1$st-order as,
\begin{align}
    f = g + Kn f_1 = g - \tau (\frac{\partial{g}}{\partial t} + u \frac{\partial g}{\partial x} + v \frac{\partial g}{\partial y} + w \frac{\partial g}{\partial z}) .
    \label{ce_expansion}
\end{align}
For the Burnett and super-Burnett equations, the above expansion can be naturally extended \cite{ohwada2004kinetic}, such as $f = g + Kn f_1 + Kn^2 f_2 + Kn^3 f_3 + \cdots$. For the above Navier-Stokes solutions, the GKS based on the kinetic BGK model has been well developed \cite{xu2001gas}. In order to simulate the flow with any realistic Prandtl number, a modification of the heat flux in the energy transport is used in this scheme, which is also implemented in the present study.

\subsection{Non-equilibrium translational and rotational temperature model}
A single temperature is assumed for translational and rotational modes in the above Navier-Stokes equations. However, it loses accuracy in the simulation of non-equilibrium flow because of the different temperatures for the translational and rotational energy modes. In the following section, an extended BGK model for non-equilibrium rotational energy is constructed and for the 1st time the corresponding three-dimensional macroscopic governing equations are derived.

For non-equilibrium multi-temperature diatomic gas flow, the above-mentioned BGK model can be extended in the following form,
\begin{align}
	\frac{\partial f}{\partial t} + + u \frac{\partial f}{\partial x} + v \frac{\partial f}{\partial y}  + w \frac{\partial f}{\partial z} = \frac{f^{eq} - f}{\tau} + \frac{g - f^{eq}}{Zr\tau} = \frac{f^{eq} - f}{\tau} + Q_s,
	 \label{boltamann_bgk_neq}
\end{align}
where an intermediate equilibrium state $f^{eq}$ different with Eq.(\ref{maxwellian}) is introduced with two temperatures, one for translational temperature and the other for rotational temperature,
\begin{align}
	f^{eq} = \rho (\frac{\lambda_t}{\pi})^{3/2} (\frac{\lambda_r}{\pi}) e^{- \lambda_t[(u - U)^2 + (v - V)^2 + (w - W)^2] - \lambda_r \xi_r^2} ,
	\label{intermediate_f}
\end{align}
where $\lambda_t = m/2kT_t$ is related to the translational temperature $T_t$, and $\lambda_r = m/2kT_r$ accounts for the rotational temperature $T_r$. Therefore, the right hand side collision operator contains two terms corresponding to the elastic and inelastic collisions respectively. Where the relaxation process becomes $f \to f^{eq} \to g$, and the inelastic collision process from $f^{eq}$ to $g$ takes a much longer time $Z_r \tau$ than that of elastic collision process by $\tau$. The additional term $ Q_s$ in the collision part accounts for the energy exchange between the translational and rotational energy, which contributes to the source term for the corresponding three-dimensional macroscopic flow evolution. The above three-dimensional extended BGK model is a natural extension for two-dimensional extended BGK model \cite{xu2008multiple}.

The relation between mass $\rho$, momentum$(\rho U, \rho V, \rho W)$, total energy $\rho E$, and rotational energy $\rho E_r$ with the distribution function $f$ is given by,
\begin{align}
W =
\begin{pmatrix}
\rho \\
\rho U \\
\rho V \\
\rho W \\
\rho E \\
\rho E_r
\end{pmatrix}
= \int \psi_{\alpha} f d \Xi, \ \ \alpha = 1, 2, 3, 4, 5, 6,
\end{align}
where $d \Xi = du dv dw d\xi_r$ and $\psi_{\alpha}$ is the component of the vector of collision invariants
\begin{align*} \psi = (\psi_1, \psi_2, \psi_3, \psi_4, \psi_5, \psi_6)^T = (1, u, v, w, \frac{1}{2}(u^2 + v^2 + w^2 + \xi_r^2), \frac{1}{2} \xi_r^2)^T.
\end{align*}
As a new temperature $\lambda_r$ is introduced, the constraint of rotational energy relaxation has to be imposed on the above extended kinetic model to self-consistently determine all unknowns. Since only mass, momentum and total energy are conserved during particle collisions, the compatibility condition for the collision term turns into,
\begin{align}
	\int (\frac{f^{eq} - f}{\tau} +  Q_s) \psi_{\alpha} d \Xi = \textbf{S} = (0,0,0,0,0,s)^T, \ \alpha = 1,2,3,4,5,6.
\end{align}
The source term for the rotational energy is from the energy exchange between translational and rotational ones during inelastic collision. The source term for the rotational energy is modeled through the Landau-Teller-Jeans-type relaxation model,
\begin{align}
	s =  \frac{(\rho E_r)^{eq} - \rho E_r}{Z_r \tau}.
\end{align}

\newpage

The equilibrium energy $(\rho E_r)^{eq}$ is determined by the assumption $T_r = T_t =T$, such that
\begin{align*}
	(\rho E_r)^{eq} = \frac{\rho}{2 \lambda_r^{eq}} \ \ \text{and} \ \ \lambda_r^{eq} = \frac{K + 3}{4} \frac{\rho}{\rho E - \frac{1}{2} \rho (U^2 + V^2 + W^2)}.
\end{align*}
Here, the collision number $Zr$ is related to the ratio of elastic collision frequency to inelastic frequency.  The particle collision time multiplied by a rotational collision number $Z_r$ models the relaxation process for the rotational energy to equilibrate with the translational one. The value $Z_r$ used in current study is given by,
\begin{align*}
    Z_r = \frac{Z_r^{\infty}}{1 + (\pi^{3/2}/2) \sqrt{T^{\ast}/T} + (\pi + \pi^2/4)(T^{\ast}/T)},
\end{align*}
where the quantity $T^{\ast}$ is the characteristic temperature of intermolecular potential, and $Z_r^{\infty}$ is the limiting value. Over a temperature range from $30K$ to $3000K$ for Nitrogen, the values $Z_r^{\infty} = 23.0 $ and $T^{\ast} = 91.5K$ are used. The local temperature $T$ in the above equation is the translational temperature. More advanced models for the energy relaxtion are discussed in \cite{parker1959rotational, koura1992statistical}.
\\
Using the intermediate state give by Eq.(\ref{intermediate_f}), with the frozen of rotational energy exchange the 1st-order Champan-Enskog expansion gives,
\begin{align}
    f = f^{eq} + Kn f_1 = f^{eq} - \tau (\frac{\partial{f^{eq}}}{\partial t} + u \frac{\partial f^{eq}}{\partial x} + v \frac{\partial f^{eq}}{\partial y} + w \frac{\partial f^{eq}}{\partial z}).
    \label{ce_expansion_neq}
\end{align}
The corresponding macroscopic non-equilibrium multi-temperature continuum equations in three-dimensions can be derived as the appendix, and the final form is given by,
\begin{align}
    \frac{\partial W}{\partial t} + \frac{\partial F}{\partial x} + \frac{\partial G}{\partial y} + \frac{\partial H}{\partial z} = \frac{\partial F_v}{\partial x} + \frac{\partial G_v}{\partial y} + \frac{\partial H_v}{\partial z}+ \textbf{S},
    \label{macro_neq}
\end{align}
with
\begin{align*}
W =
\begin{pmatrix}
\rho \\
\rho U \\
\rho V \\
\rho W \\
\rho E \\
\rho E_r
\end{pmatrix} \ \
F =
\begin{pmatrix}
\rho U\\
\rho U^2 + p \\
\rho UV \\
\rho UW \\
(\rho E + p) U \\
\rho E_r U
\end{pmatrix}
G =
\begin{pmatrix}
\rho V\\
\rho UV \\
\rho V^2 + p \\
\rho VW \\
(\rho E + p) V \\
\rho E_r V
\end{pmatrix}
H =
\begin{pmatrix}
\rho W\\
\rho UW \\
\rho VW \\
\rho W^2 + p \\
(\rho E + p) W \\
\rho E_r W
\end{pmatrix},
\end{align*}
and
\begin{align*}
F_v =
\begin{pmatrix}
0\\
\tau_{xx} \\
\tau_{xy} \\
\tau_{xz} \\
U \tau_{xx}  + V \tau_{xy} + W \tau_{xz} + q_x \\
U \tau_{tr} + q_{rx}
\end{pmatrix}
G_v =
\begin{pmatrix}
0\\
\tau_{yx} \\
\tau_{yy} \\
\tau_{yz} \\
U \tau_{yx}  + V \tau_{yy} + W \tau_{yz} + q_y \\
V \tau_{tr} + q_{ry}
\end{pmatrix}
\end{align*}
\begin{align*}
H_v =
\begin{pmatrix}
	0\\
	\tau_{zx} \\
	\tau_{zy} \\
	\tau_{zz} \\
	U \tau_{zx}  + V \tau_{zy} + W \tau_{zz} + q_z \\
	W \tau_{tr} + q_{rz}
\end{pmatrix},
\end{align*}
where $\rho E = \frac{1}{2}\rho(\textbf{U}^2 + 3RT_t + KRT_r)$ is the total energy, and $\rho E_r =  \rho  R T_r$ with $K = 2$ is the rotational energy. The pressure $p$ is related to the translational  temperature as $p = \rho R T_t$. Meanwhile, the viscous normal stress terms are
\begin{align*}
    \tau_{xx} &= \tau p[2 \frac{\partial U}{\partial x} - \frac{2}{3}(\frac{\partial U}{\partial x} + \frac{\partial V}{\partial y} + \frac{\partial W}{\partial z})] - \frac{\rho K}{2(K + 3)} \frac{1}{Zr} (\frac{1}{\lambda_t} - \frac{1}{\lambda_r}), \\
    \tau_{yy} &= \tau p[2 \frac{\partial V}{\partial y} - \frac{2}{3}(\frac{\partial U}{\partial x} + \frac{\partial V}{\partial y} + \frac{\partial W}{\partial z})] - \frac{\rho K}{2(K + 3)} \frac{1}{Zr} (\frac{1}{\lambda_t} - \frac{1}{\lambda_r}), \\
    \tau_{zz} &= \tau p[2 \frac{\partial W}{\partial z} - \frac{2}{3}(\frac{\partial U}{\partial x} + \frac{\partial V}{\partial y} + \frac{\partial W}{\partial z})] - \frac{\rho K}{2(K + 3)} \frac{1}{Zr} (\frac{1}{\lambda_t} - \frac{1}{\lambda_r}),
\end{align*}
with viscous shear stress term given by,
\begin{align*}
    \tau_{xy} = \tau_{yx} = \tau p (\frac{\partial U}{\partial y} + \frac{\partial V}{\partial x}), \\
    \tau_{xz} = \tau_{zx} = \tau p (\frac{\partial U}{\partial z} + \frac{\partial W}{\partial x}), \\
    \tau_{yz} = \tau_{zy} = \tau p (\frac{\partial V}{\partial z} + \frac{\partial W}{\partial y}),
\end{align*}
and heat conduction terms are
\begin{align*}
    q_x = \tau p [\frac{K}{4} \frac{\partial}{\partial x} (\frac{1}{\lambda_r}) + \frac{5}{4} \frac{\partial}{\partial x} (\frac{1}{\lambda_t})], \\
    q_y = \tau p [\frac{K}{4} \frac{\partial}{\partial y} (\frac{1}{\lambda_r}) + \frac{5}{4} \frac{\partial}{\partial y} (\frac{1}{\lambda_t})], \\
    q_z = \tau p [\frac{K}{4} \frac{\partial}{\partial z} (\frac{1}{\lambda_r}) + \frac{5}{4} \frac{\partial}{\partial z} (\frac{1}{\lambda_t})].
\end{align*}
The following terms are related to governing equation of rotational energy $\rho E_r$ as,
\begin{align*}
    \tau_{rt} &= \frac{3 \rho K}{4 (K + 3)} \frac{1}{Zr} (\frac{1}{\lambda_t} - \frac{1}{\lambda_r}), \\
    q_{rx}    &= \tau p \frac{K}{4} \frac{\partial}{\partial x} \frac{1}{\lambda_r}, \\
    q_{ry}    &= \tau p \frac{K}{4} \frac{\partial}{\partial y} \frac{1}{\lambda_r}, \\
    q_{rz}    &= \tau p \frac{K}{4} \frac{\partial}{\partial z} \frac{1}{\lambda_r}.
\end{align*}
The source term in Eq.(\ref{macro_neq}) is given by,
\begin{align*}
	\textbf{S} = (0, 0, 0, 0, 0,  \frac{(\rho E_r)^{eq} - \rho E_r}{Z_r \tau}).
\end{align*}

Instead of the bulk viscosity term in the standard NS equations, a relaxation term between translational and rotational energy is obtained in the above equations to model the non-equilibrium process. The bulk viscosity term in NS equations,
\begin{equation*}
\begin{aligned}
    \frac{2}{3}\frac{K}{K + 3} \tau p (U_x + V_y + W_z),
\end{aligned}
\end{equation*}
is replaced by the temperature relaxation term in the above Eq.(\ref{macro_neq}),
\begin{equation*}
\begin{aligned}
	- \frac{\rho K}{2(K + 3)} \frac{1}{Zr} (\frac{1}{\lambda_t} - \frac{1}{\lambda_r}) = \frac{\rho R}{Z_r} \frac{K}{K + 3} (T_r -T_t).
\end{aligned}
\end{equation*}
In the limiting case of small departures from equilibrium, the rotational energy equation becomes
\begin{equation*}
\begin{aligned}
    (\rho E_r)_t + (\rho E_r U)_x + (\rho E_r V)_y + (\rho E_r W)_z = \frac{\rho R}{Z_r \tau} \frac{3}{K + 3} (T_t - T_r),
\end{aligned}
\end{equation*}
and with the Euler approximation for the right hand side of the above equation, we have
\begin{equation*}
\begin{aligned}
    \label{del_rot_trans}
    T_t - T_r = -\frac{2}{3} Z_r \tau T (U_x + V_y + W_z).
\end{aligned}
\end{equation*}
Based on above equation, the normal bulk viscosity term can be exactly recovered, given by
\begin{equation*}
\begin{aligned}
    \frac{2}{3}\frac{K}{K + 3} \tau p (U_x + V_y + W_z) = \frac{\rho R}{Z_r} \frac{K}{K + 3} (T_r -T_t).
\end{aligned}
\end{equation*}

With the above macroscopic modeling equations for a multi-temperature system, the non-equilibrium flow in the near continuum regime is modeled beyond the NS assumption. The bulk viscosity is replaced by a relaxation term between translational and rotational energy, which seems more physically meaningful than the bulk viscosity assumption \cite{xu2006continuum, xu2008multiple}, for the flows inside the shock layer or the hypersonic flow near isothermal boundary. However, we are supposed to keep in mind that the extended kinetic equation Eq.(\ref{boltamann_bgk_neq}) will be directly used in the numerical scheme in the following part, instead of solving the nonlinear system Eq.(\ref{macro_neq}).

\newpage

\section{High-order finite volume non-equilibrium gas-kinetic scheme}
The extended model proposed in the previous section is solved based on the conservative finite volume method GKS \cite{xu2001gas}. The numerical fluxes at cell interfaces are evaluated based on the general time-dependent gas distribution solution. In this paper, a high-order non-equilibrium finite volume GKS will be constructed, where the additional source term is dealt with fourth-order Simpson interpolation rule.

\subsection{Three-dimensional finite volume scheme}

Taking moments of Eq.(\ref{boltamann_bgk_neq}) and integrating over the control volume $V_{ijk} = \overline{x_i} \times \overline{y_j} \times \overline{z_k}$ with $\overline{x_i} = [x_i - \frac{\Delta x}{2}, x_i + \frac{\Delta x}{2}]$, $\overline{y_j} = [y_j - \frac{\Delta y}{2}, y_j + \frac{\Delta y}{2}]$, $\overline{z_k} = [z_k - \frac{\Delta z}{2}, z_k + \frac{\Delta z}{2}]$, the three-dimensional non-equilibrium finite volume scheme can be written as
\begin{equation}
\begin{aligned}
    &\frac{d W_{ijk}}{d t} = L(W_{ijk}) = \frac{1}{|V_{ijk}|} [\int_{\overline{y_j} \times \overline{z_k}} (F_{i - 1/2, j, k} - F_{i + 1/2, j, k}) dy dz \\
    &+ \int_{\overline{x_i} \times \overline{z_k}} (G_{i, j - 1/2, k} - G_{i, j + 1/2, k}) dx dz + \int_{\overline{x_i} \times \overline{y_j}} (H_{i, j, k - 1/2} - G_{i, j, k + 1/2}) dx dy] \\
    &+ 	S_{ijk},
\end{aligned}
\end{equation}
where $W_{ijk}$ is the cell averaged flow variables of mass, momentum, total energy, and rotational energy, and $S_{ijk}$ is cell averaged source term for the rotational energy. All of them are averaged over control volume $V_{ijk}$ and the volume of the numerical cell is $|V_{ijk}| = \Delta x \Delta y \Delta z$. Here, numerical fluxes in $x-\text{direction}$ is presented as an example
\begin{align}
    \int_{\overline{y_j} \times \overline{z_k}}  F_{i + 1/2, j, k} dy dz = F_{\textbf{x}_{i + 1/2, j, k}, t} \Delta y \Delta z.
\end{align}
Based on the fifth-order weighted essentially non-oscillatory scheme (WENO-JS) \cite{jiang1996efficient} for the spatial reconstruction on the primitive flow variables, the reconstructed pointwise values and the spatial derivatives in normal and tangential direction can be obtained. In the smooth flow computation, the linear form of WENO-JS is adopted to reduce the dissipation. The numerical fluxes $F_{\textbf{x}_{i + 1/2, j, k}, t}$ can be provided by the flow solvers, which can be evaluated by taking moments of the gas distribution function as

\newpage

\begin{align}
    F_{\textbf{x}_{i + 1/2, j, k}, t}= \int \psi_{\alpha} u f(\textbf{x}_{i + 1/2, j, k}, t, \textbf{u}, \xi) d \Xi, \ \ \alpha = 1, 2, 3, 4, 5, 6,
    \label{flux_integration}
\end{align}
where $f(\textbf{x}_{i + 1/2, j, k}, t, \textbf{u}, \xi)$ is based on the integral solution of BGK equation Eq.(\ref{boltamann_bgk_neq}) at the cell interface
\begin{align}
f(\textbf{x}_{i + 1/2, j, k}, t, \textbf{u}, \xi_r) = \frac{1}{\tau} \int_0^t f^{eq}(\textbf{x}', t', \textbf{u}, \xi_r) e^{-(t - t')/\tau} d t' + e^{-t/\tau} f_0(-\textbf{u}t, \xi_r),
\end{align}
where $\textbf{x}_{i + 1/2, j, k} = \textbf{0}$ is the location of the cell interface, $\textbf{u} = (u, v, w)$ is the particle velocity, $\textbf{x}_{i + 1/2, j, k} = \textbf{x}' + \textbf{u} (t - t')$ is the trajectory of particles.  $f_0$ is the initial gas distribution, and $f^{eq}$ is the corresponding intermediate equilibrium state as Eq.(\ref{intermediate_f}). $f^{eq}$ and $f_0$ can be constructed as
\begin{equation*}
\begin{aligned}
    	f^{eq} = f^{eq}_0(1 + \overline{a} x + \overline{b} y + \overline{c} z + \overline{A} t),
\end{aligned}
\end{equation*}
and
\begin{equation*}
\begin{aligned}
f_0 =
\begin{cases}
f^{eq}_l [1 +  (a_l x + b_l y + c_l z) - \tau (a_l u + b_l v + c_l w + A_l)], &x \leq 0, \\
f^{eq}_r [1 +  (a_r x + b_r y + c_r z) - \tau (a_r u + b_r v + c_r w + A_r)], &x > 0,
\end{cases}
\end{aligned}
\end{equation*}
where $f^{eq}_l$ and $f^{eq}_r$ are the initial gas distribution functions on both sides of a cell interface.$f^{eq}_0$ is the initial equilibrium state located at cell interface, which can be determined through the compatibility condition
\begin{align*}
	\int \psi_{\alpha} f^{eq}_0 d \Xi = \int_{u>0} \psi_{\alpha} f^{eq}_l d \Xi + \int_{u<0} \psi_{\alpha} f^{eq}_r d \Xi, \ \ \alpha = 1, 2, 3, 4, 5, 6.
\end{align*}
For a second-order flux, the time-dependent gas distribution function at the cell interfaces is evaluated as
\begin{equation}
\begin{aligned}
    \label{formalsolution_neq}
    f(\textbf{x}_{i + 1/2, j, k}, t, \textbf{u}, \xi_r) &= (1 - e^{-t/\tau}) f^{eq}_0 + ((t + \tau) e^{-t\tau} - \tau) (\overline{a} u + \overline{b} v + \overline{c} w) f^{eq}_0   \\
    &+ (t - \tau + \tau e^{-t\tau}) \overline{A} f^{eq}_0  \\
    &+ e^{-t/\tau} f^{eq}_l [1 - (\tau + t) (a_l u + b_l v + c_l w) - \tau A_l] (1 - H(u)) \\
    &+ e^{-t/\tau} f^{eq}_r [1 - (\tau + t) (a_r u + b_r v + c_r w) - \tau A_r] H(u),
\end{aligned}
\end{equation}
where the coefficients in Eq.(\ref{formalsolution_neq}) can be determined by the spatial derivatives of macroscopic flow variables and the compatibility condition.
For three-dimensional diatomic gas, the expansion of spatial variation $\partial f^{eq}/\partial x$ is given by,
\begin{align}
    \frac{\partial f^{eq}}{\partial x} = \frac{1}{\rho} (a_1 + a_2 u + a_3 v + a_4 w + a_5(u^2 + v^2 + w^2) + a_6 \xi_r^2) f^{eq} = \frac{1}{\rho} a f^{eq},
    \label{micro_coefficients}
\end{align}
where all the coefficients in Eq.(\ref{micro_coefficients}) can be explicitly determined by the relation with the microscopic and macroscopic variables at the cell interface, i.e., $W = \int \psi_{\alpha} f^{eq} du dv dw d\xi_r$ and $\partial W/\partial x = (1/\rho) \int \psi_{\alpha} a f^{eq} du dv dw d\xi_r$, where $W = (\rho, \rho U, \rho V, \rho W, \rho E, \rho E_r)$ are the flow variables. The components of coefficients $a$ in Eq.(\ref{micro_coefficients}) can be expressed as
\begin{align*}
a_6 &= 2 \frac{\lambda_r^2}{K} (2 \frac{\partial (\rho E_r)}{\partial x} - \frac{1}{2}\frac{K}{\lambda_r} \frac{\partial \rho}{\partial x}), \\
a_5 &= \frac{2 \lambda_t^2}{3}(B - 2 U A_1 - 2V A_2 - 2W A_3), \\
a_4 &= 2 \lambda_t A_3 - 2W a_5, \\
a_3 &= 2 \lambda_t A_2 - 2V a_5, \\
a_2 &= 2 \lambda_t A_1 - 2U a_5, \\
a_1 &= \frac{\partial \rho}{\partial x} - a_2 U - a_3 V - a_4 W - a_5(U^2 + V^2 + W^2 + \frac{3}{\lambda_t}) - a_6 \frac{K}{2 \lambda_r},
\end{align*}
with the defined variables
\begin{align*}
B &= 2 \frac{\partial (\rho E - \rho E_r)}{\partial x} - (U^2 + V^2 + W^2 + \frac{3}{\lambda_t})\frac{\partial \rho}{\partial x}, \\
A_1 &= \frac{\partial (\rho U)}{\partial x} - U \frac{\partial \rho}{\partial x}, \\
A_2 &= \frac{\partial (\rho V)}{\partial y} - V \frac{\partial \rho}{\partial x}, \\
A_3 &= \frac{\partial (\rho W)}{\partial z} - W \frac{\partial \rho}{\partial x}.
\end{align*}
In a similar way, the temporal variation of $\partial f^{eq}/\partial t$ can be expanded and the corresponding coefficients can be obtained from the compatibility condition for the Chapman-Enskog expansion
\begin{align*}
\int \psi_{\alpha} (\frac{\partial{f^{eq}}}{\partial t} + u \frac{\partial f^{eq}}{\partial x} + v \frac{\partial f^{eq}}{\partial y} + w \frac{\partial f^{eq}}{\partial z}) d \Xi = 0,
\end{align*}
where the above six equations uniquely determine six unknowns in $A$, i.e., $A = A_1 + A_2 u + A_3 v + A_4 w + A_5(u^2 + v^2 + w^2) + A_6 \xi_r^2$.

Here, the second-order accuracy in time can be achieved by one step integration, with the second-order gas-kinetic flux solver Eq.(\ref{formalsolution_neq}). Based on a higher-order expansion of the equilibrium state around a cell interface, the one-stage high-order GKS has been developed successfully \cite{li2010high}. However, the one-stage gas-kinetic solver become very complicated, especially for three-dimensional multidimensional computations. In order to reduce the complexity of high-order scheme, the technique of a two-stage fourth-order method will be used here for the development of high-order scheme for non-equilibrium flow.

\subsection{Two-stage high-order temporal discretization}
In recent study, a two-stage fourth-order time-accurate discretization was developed for Lax-Wendroff flow solvers, particularly applied for hyperbolic equations with the generalized Riemann problem (GRP) solver \cite{li2016two} and the GKS \cite{pan2016efficient}. Such method provides a reliable framework to develop a high-order three-dimensional non-equilibrium GKS with a second-order flux function Eq.(\ref{formalsolution_neq}) only, where the source terms will be treated by high-order interpolation. Key point for this two-stage high-order method is to use the time derivative of a flux function. In order to obtain the time derivative of flux function at $t_n$ and $t_{\ast} = t_n + \Delta t/2$, the flux function should be approximated as a linear function of time within a time interval.

According to the numerical fluxes at cell interface Eq.(\ref{flux_integration}), the following notation is introduced
\begin{align}
    \mathbb{F}_{i + 1/2, j, k}(W^n, \delta) &= \int_{t_n}^{t_n + \delta} \mathbf{F}_{i + 1/2, j, k}(W^n, t) dt = \int_{t_n}^{t_n + \delta} F_{\textbf{x}_{i + 1/2, j, k}, t} dt.
    \label{Flux_twostage}
\end{align}
In the time interval $[ t_n, t_n + \Delta t/2]$, the flux is expanded as the following linear form
\begin{align}
    \mathbf{F}_{i + 1/2, j, k} (W^n, t) = \mathbf{F}_{i + 1/2, j, k}^n +　\partial_t \mathbf{F}_{i + 1/2, j, k}^n (t - t_n).
    \label{linear_flux}
\end{align}
Based on Eq.(\ref{Flux_twostage}) and linear expansion of flux as Eq.(\ref{linear_flux}), the coefficients $\mathbf{F}_{i + 1/2, j, k}(W^n, t_n)$ and $\partial_t \mathbf{F}_{i + 1/2, j, k}(W^n, t_n)$ can be determined as,
\begin{align*}
    \mathbf{F}_{i + 1/2, j, k}(W^n, t_n) \Delta t + \frac{1}{2} \partial_t \mathbf{F}_{i + 1/2, j, k}(W^n, t_n)  \Delta t^2 &= 	\mathbb{F}_{i + 1/2, j, k}(W^n, \Delta t), \\
    \frac{1}{2} \mathbf{F}_{i + 1/2, j, k}(W^n, t_n) \Delta t + \frac{1}{8} \partial_t \mathbf{F}_{i + 1/2, j, k}(W^n, t_n)  \Delta t^2 &= \mathbb{F}_{i + 1/2, j, k}(W^n, \Delta t/2).
\end{align*}
By solving the linear system, we have
\begin{equation}
\begin{aligned}
    \label{flux_der}
    \mathbf{F}_{i + 1/2, j, k}(W^n, t_n) &= (4 \mathbb{F}_{i + 1/2, j, k}(W^n, \Delta t/2) - \mathbb{F}_{i + 1/2, j, k}(W^n, \Delta t))/\Delta t,  \\
    \partial_t \mathbf{F}_{i + 1/2, j, k}(W^n, t_n) &= 4(\mathbb{F}_{i + 1/2, j, k}(W^n, \Delta t) - \mathbb{F}_{i + 1/2, j, k}(W^n, \Delta t/2))/\Delta t^2 ,
\end{aligned}
\end{equation}
and $\mathbf{F}_{i + 1/2, j, k}(W^{\ast}, t_{\ast})$,$\partial_t \mathbf{F}_{i + 1/2, j, k}(W^{\ast}, t_{\ast})$ for the intermediate state $t_{\ast}$ can be constructed similarly.

With these notations, the three-dimensional high-order non-equilibrium algorithm for multi-temperature flow is given by \\
(i) With the initial reconstruction, update $W^{\ast}$ at $t_{\ast} = t_n + \Delta t/2$ by
\begin{equation}
\begin{aligned}
    \label{qq_star}
    W_{ijk}^{\ast} = W_{ijk}^{n} &- \frac{1}{\Delta x} [\mathbb{F}_{i + 1/2, j, k}(W^n, \Delta t/2) - \mathbb{F}_{i - 1/2, j, k}(W^n, \Delta t/2)]  \\
    &- \frac{1}{\Delta y} [\mathbb{G}_{i, j + 1/2, k}(W^n, \Delta t/2) - \mathbb{G}_{i, j - 1/2, k}(W^n, \Delta t/2)] \\
    &- \frac{1}{\Delta z} [\mathbb{H}_{i, j, k + 1/2}(W^n, \Delta t/2) - \mathbb{H}_{i, j, k - 1/2}(W^n, \Delta t/2)]  \\
    &+ S_{ijk}^{\ast} \frac{\Delta t}{2},
\end{aligned}
\end{equation}
and compute the fluxes and their derivatives by Eq.(\ref{flux_der}) for future use,
\begin{align*}
    \mathbf{F}_{i + 1/2, j, k}(W^n, t_n), \ &\mathbf{G}_{i, j + 1/2, k}(W^n, t_n), \ \mathbf{H}_{i, j, k + 1/2}(W^n, t_n), \\
    \partial_t \mathbf{F}_{i + 1/2, j, k}(W^n, t_n), \ &\partial_t \mathbf{G}_{i, j + 1/2, k}(W^n, t_n), \ \partial_t \mathbf{H}_{i, j, k + 1/2}(W^n, t_n).
\end{align*}

\newpage

(ii) Reconstruct intermediate value $W_{ijk}^{\ast}$ and compute
\begin{align*}
    \partial_t \mathbf{F}_{i + 1/2, j, k}(W^{\ast}, t_{\ast}), \ &\partial_t \mathbf{G}_{i, j + 1/2, k}(W^{\ast}, t_{\ast}),\ \partial_t \mathbf{H}_{i, j, k + 1/2}(W^{\ast}, t_{\ast}),
\end{align*}
where the derivatives are determined by Eq.(\ref{flux_der}) in the time interval $[t_{\ast}, t_{\ast} + \Delta t]$. \\
(iii) Update $W_{ijk}^{n + 1}$ by
\begin{equation}
\begin{aligned}
    \label{qq_nn}
    W_{ijk}^{n + 1} = &W_{ijk}^{n} - \frac{\Delta t}{\Delta x}[\mathscr{F}^n_{i + 1/2, j, k} - \mathscr{F}^n_{i - 1/2, j, k}]  \\
    &- \frac{\Delta t}{\Delta y}[\mathscr{G}^n_{i, j + 1/2, k} - \mathscr{G}^n_{i, j - 1/2, k}] - \frac{\Delta t}{\Delta z}[\mathscr{H}^n_{i, j, k + 1/2} - \mathscr{H}^n_{i, j, k - 1/2}] \\
    &+ S_{ijk}^{n + 1} \Delta t,
\end{aligned}
\end{equation}
where $\mathscr{F}^n_{i + 1/2, j, k}$, $\mathscr{G}^n_{i, j + 1/2, k}$ and $\mathscr{H}^n_{i, j, k + 1/2}$ are the numerical fluxes and expressed as
\begin{align*}
\mathscr{F}^n_{i + 1/2, j, k} &= \mathbf{F}_{i + 1/2, j, k}(W^n, t_n) + \frac{\Delta t}{6} [\partial_t \mathbf{F}_{i + 1/2, j, k}(W^n, t_n) + 2 \partial_t \mathbf{F}_{i + 1/2, j, k}(W^{\ast}, t_{\ast})], \\
\mathscr{G}^n_{i, j + 1/2, k} &= \mathbf{G}_{i, j + 1/2, k}(W^n, t_n) + \frac{\Delta t}{6} [\partial_t \mathbf{G}_{i, j + 1/2, k}(W^n, t_n) + 2 \partial_t \mathbf{G}_{i, j + 1/2, k}(W^{\ast}, t_{\ast})], \\
\mathscr{H}^n_{i, j, k + 1/2} &= \mathbf{H}_{i, j, k + 1/2}(W^n, t_n) + \frac{\Delta t}{6} [\partial_t \mathbf{H}_{i, j, k + 1/2}(W^n, t_n) + 2 \partial_t \mathbf{H}_{i, j, k + 1/2}(W^{\ast}, t_{\ast})],
\end{align*}
where $S_{ijk}^{\ast}$ and $S_{ijk}^{n + 1}$ are source terms, which will be solved through a high-order semi-implicit way.

\subsection{Fourth-order Simpson interpolation for source term}
Let $s_{ijk}$ denotes the source component for rotational energy $\rho E_r$, while other components in source term $S_{ijk}$ are zero. Here, $\rho E_r$ can be updated using an semi-implicit scheme based on fourth-order Simpson interpolation rule. \\
(i) Update  $(\rho E_r)^{\ast}$ at $t^{\ast} = t_n + \Delta t/2$ by
\begin{align*}
    (\rho E_r)_{ijk}^{\ast} &= (\rho E_r)_{ijk}^{n} + (RHS)^{\ast}_{ijk} + \frac{\Delta t^{\ast}}{2}(s_{ijk}^n + s_{ijk}^{n + 1}), \\
    s_{ijk}^{n    } &= \frac{(\rho E_r^{eq})_{ijk}^{n    } -  (\rho E_r)_{ijk}^{n    } }{(Z_r \tau)_{ijk}^{n    }}, \\
    s_{ijk}^{\ast}  &= \frac{(\rho E_r^{eq})_{ijk}^{\ast} -  (\rho E_r)_{ijk}^{\ast} }{(Z_r \tau)_{ijk}^{\ast}},
\end{align*}
thus
\begin{align}
    (\rho E_r)_{ijk}^{\ast} &= \frac{2(Z_r \tau)_{ijk}^{\ast} }{2(Z_r \tau)_{ijk}^{\ast} + \Delta t^{\ast}}[(\rho E_r)_{ijk}^{n} + (RHS)^{\ast}_{ijk} + \frac{\Delta t^{\ast}}{2}(s_{ijk}^n + \frac{(\rho E_r^{eq})_{ijk}^{\ast}}{(Z_r \tau)_{ijk}^{\ast}})],
    \label{source_star}
\end{align}
where $\Delta t^{\ast} = \Delta t/ 2$ and $(RHS)^{\ast}_{ijk}$ represents the component for rotational energy on the right hand side of Eq.(\ref{qq_star}) without source term. $(\rho E_r)^{\ast}$ can be updated based on Eq.(\ref{source_star}), as the right hand side terms are known after updating the flow variables through fluxes at $t^{\ast}$.\\
(ii) Update $(\rho E_r)^{n + 1}$ at $t^{n + 1}$ by
\begin{align*}
    (\rho E_r)_{ijk}^{n + 1} &= (\rho E_r)_{ijk}^{n} + (RHS)_{ijk}^{n + 1} + \frac{\Delta t}{6}(s_{ijk}^n + 4 s_{ijk}^* + s_{ijk}^{ n + 1}), \\
    s_{ijk}^{n + 1} &= \frac{(\rho E_r^{eq})_{ijk}^{n + 1} -  (\rho E_r)_{ijk}^{n + 1} }{(Z_r \tau)_{ijk}^{n + 1}},
\end{align*}
thus
\begin{align}
    (\rho E_r)_{ijk}^{n + 1} = \frac{6(Z_r \tau)_{ijk}^{n + 1} }{6(Z_r \tau)_{ijk}^{n + 1} + \Delta t}[(\rho E_r)_{ijk}^{n} + (RHS)_{ijk}^{n + 1} + \frac{\Delta t}{6}(s_{ijk}^n + 4 s_{ijk}^* +  \frac{(\rho E_r^{eq})_{ijk}^{n + 1}}{(Z_r \tau)_{ijk}^{n + 1}})],
    \label{source_plus}
\end{align}
where $(RHS)^{n+1}_{ijk}$ represents the component for rotational energy on the right hand side of Eq.(\ref{qq_nn}) without source term. The right hand side terms are in Eq.(\ref{source_plus}) are known after updating the flow variables through fluxes at $t^{n + 1}$, so  $(\rho E_r)^{n + 1}$ can be updated based on the fourth-order Simpson interpolation rule.

\section{Numerical examples}
In this section, numerical tests from  smooth flow to hypersonic ones will be presented to validate our numerical scheme. The collision time $\tau$ takes
\begin{align*}
    \tau = \frac{\mu}{p} + C \frac{|p_L - p_R|}{|p_L + p_R|} \Delta t,
\end{align*}
where $\mu$ is the viscous coefficient obtained from Sutherland's Law, and $C$ is set to $1.5$ in the computation. $p_L$ and $p_R$ denotes the pressure on the left and right hand sides at the cell interface, which will reduce to $\tau = \mu /p$ in the smooth flow region. $\Delta t$ is the time step which is determined according to the CFL number, which takes $0.3$ in these computations.

\subsection{Decaying homogeneous isotropic turbulence}
Decaying homogeneous isotropic turbulence (DHIT) provides a benchmark for testing the dissipative behavior of numerical scheme. In current study, the reference experiment is conducted by Comte-Bellot et al. \cite{comte1971simple}, with Taylor Reynolds number $Re_{\lambda} = 71.6$ and turbulent Mach number $Ma_t = 0.2$.  Here, computation domain is $(2\pi)^3$ box with $128^3$ uniform grids. Vremann-type large dddy simulation (LES) model \cite{vreman2004eddy} is implemented with periodic boundary condition in 6 faces.

The turbulent fluctuating velocity as $u^{'}$, the Taylor microscale $\lambda$, the Taylor Reynolds number $Re_{\lambda}$ and the turbulent Mach number $Ma_t$ are defined as
\begin{align*}
    u^{'} &= <(u_1^2 + u_2^2 + u_3^2)/3>^{1/2}, \\
    \lambda^2 &= \frac{{u^{'}}^2}{<(\partial u_1/ \partial x_1)^2>}, \\
    Re_{\lambda} &= \frac{u^{'} \lambda}{\nu}, \\
    Ma_t &= \frac{<u_1^2 + u_2^2 + u_3^2>^{1/2}}{c},
\end{align*}
where $<\cdots>$ represents the space average in computation domain. $c$ represents the local sound speed, and $\nu$ represents the kinematic viscosity coefficient as $\mu/\rho$. The initial velocity fields is computed from experiments energy spectral, with constant pressure, density and temperature. For multi-temperature simulation, collision number $Z_r = 5$ is used. The rotational temperature is initiated with the same value as translational temperature.
\begin{figure}[htp]
	\centering
	\includegraphics[height=0.45\textwidth]{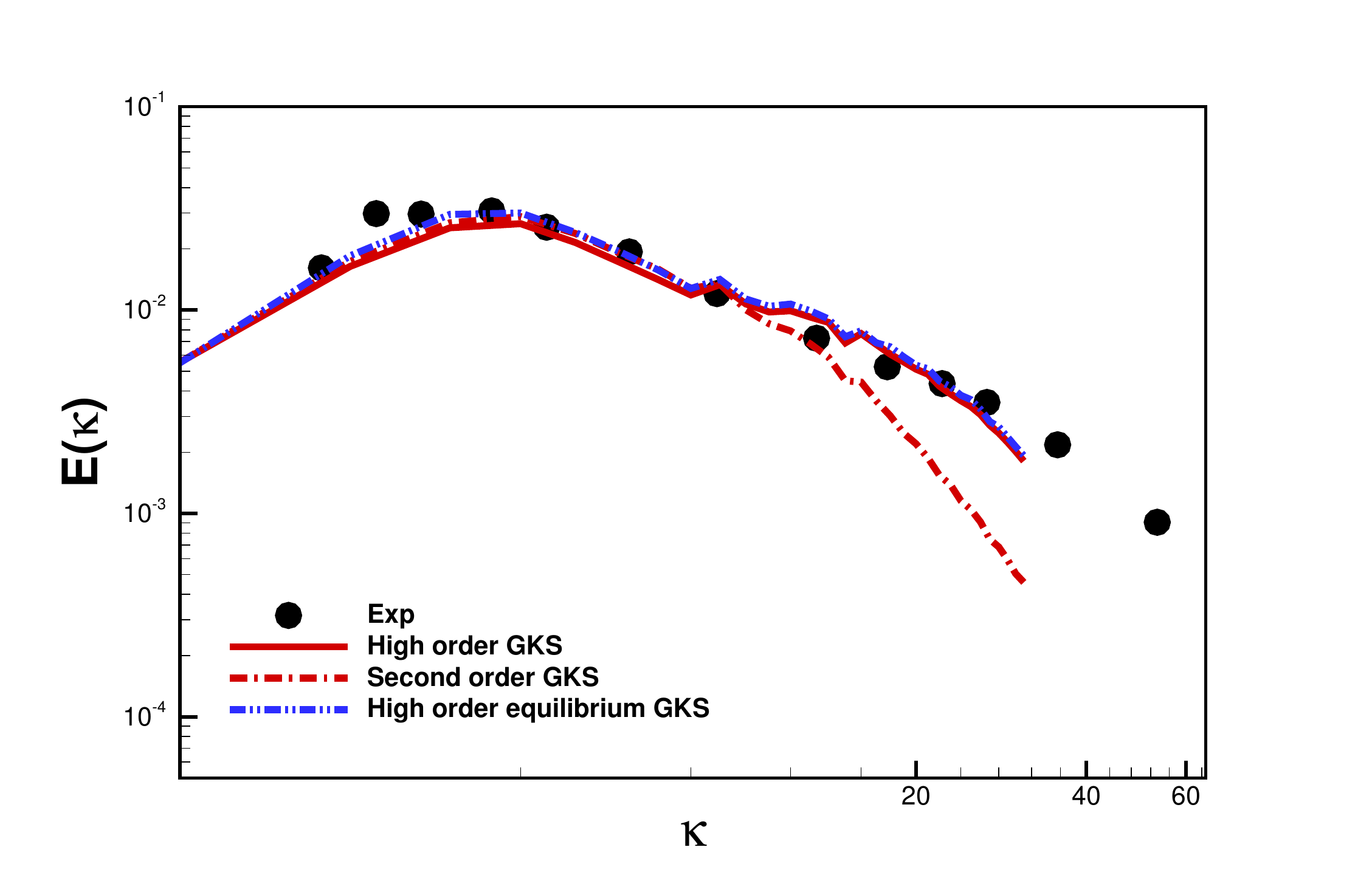}
	\caption{Comparison of TKE spectral on high order equilibrium GKS, high order GKS and second order GKS with collision number $Z_r = 5$ at dimensionless time $t^{\ast}=0.87$. The experimental data is from \cite{comte1971simple}.} 
	\label{dhit_spectral}
\end{figure}
The following quantities of turbulence have been computed in our simulations
\begin{align*}
E(\kappa) &= \frac{1}{2}\int_{\kappa_{min}}^{\kappa_{max}} \Phi_{ii}(\bm{\kappa}) \delta(|\bm{\kappa}| - \kappa) d \bm{\kappa},\\
M_{loc} &= \frac{(u_1^2 + u_2^2 + u_3^2)^{1/2}}{c},\\
\Delta_T &= \frac{T_{rn} - R_{ot}}{T_0},
\end{align*}
where velocity spectral $\Phi_{ii}$ is the Fourier transform of two-point correlation, with wave number $\kappa_{min} = 0$ and $\kappa_{max} = 64$. $T_0$ is the initial temperature, while $T_{rn}$ and $R_{ot}$ represent the translational temperature and rotational temperature, respectively.

\begin{figure}[htp]
	\begin{minipage}{0.48\linewidth}
		\includegraphics[height=0.9\textwidth]{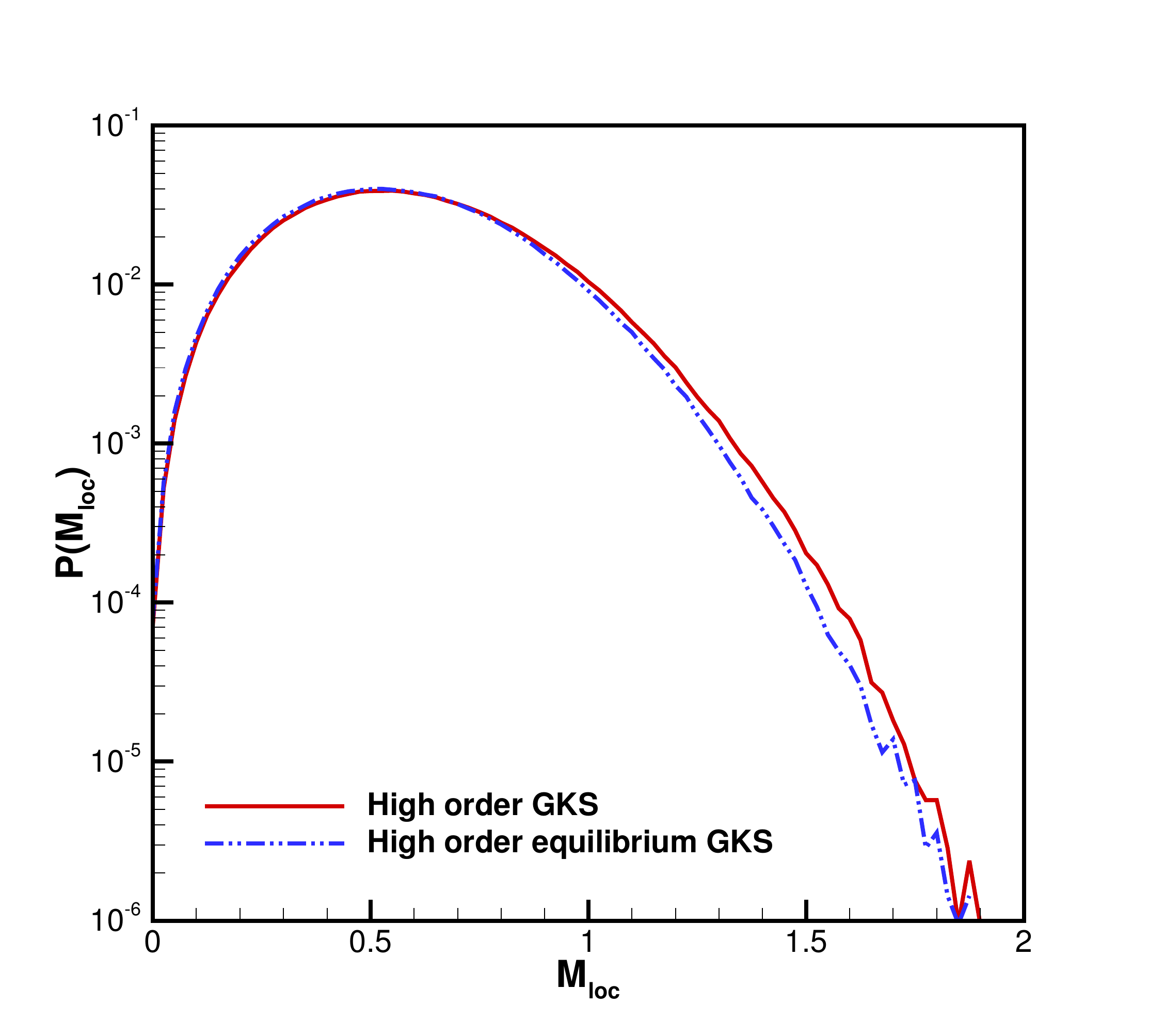}
	\end{minipage}
	\begin{minipage}{0.48\linewidth}
		\includegraphics[height=0.9\textwidth]{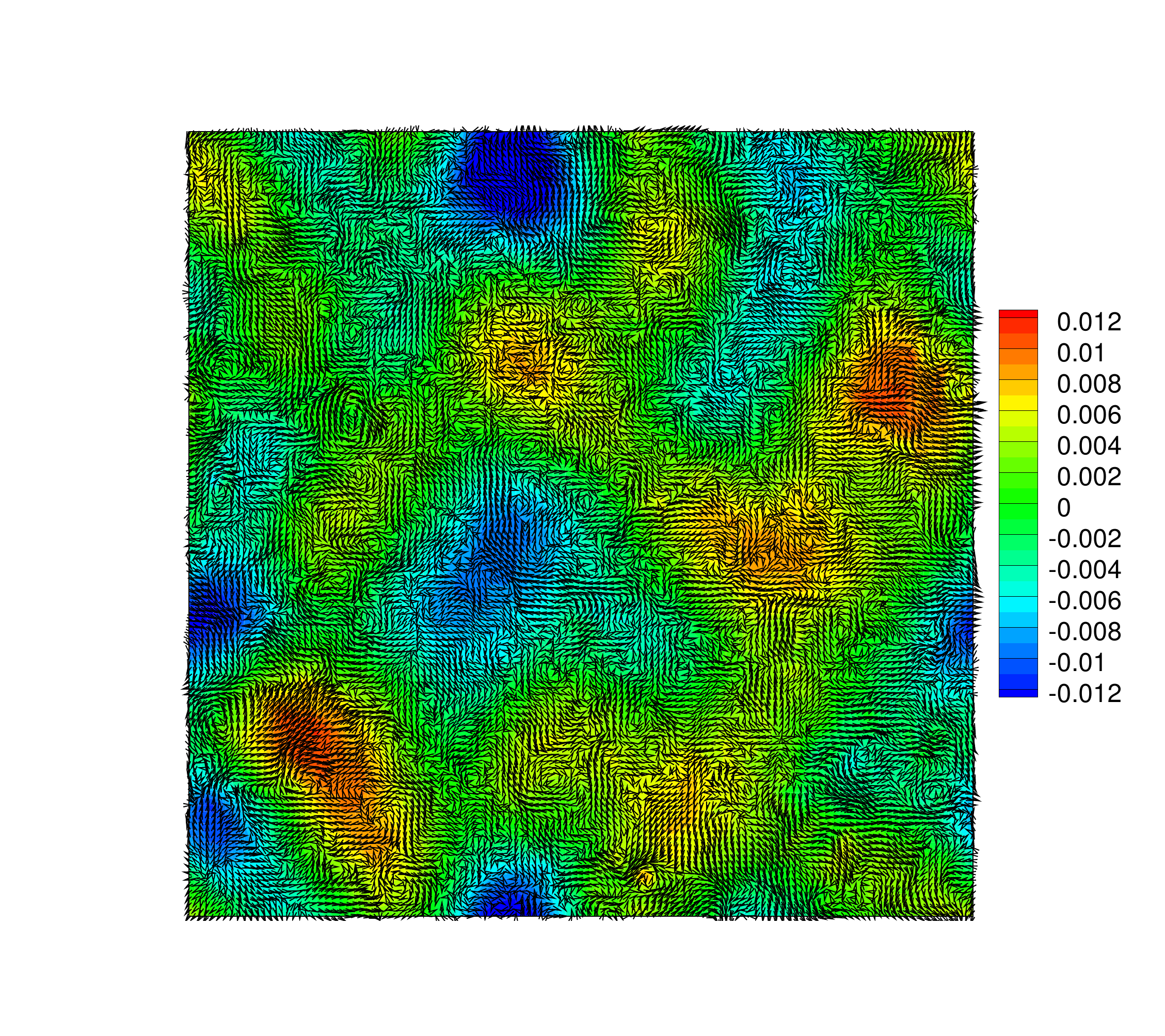}
	\end{minipage}
	\caption{PDF of the local Mach number $M_{loc}$ (left) and contour of $\Delta_T$ (right ) on the $z = 0.5$ plane at dimensionless time $t^{\ast}=0.87$ .} 
	\label{dhit_pdf_dt}
\end{figure}

Figure \ref{dhit_spectral} shows the turbulence kinetic energy (TKE) spectral at dimensionless time $t^{\ast}=0.87$, based on high order equilibrium GKS, high order GKS and second order GKS. Without special statement, high order GKS denotes current high order non-equilibrium multi-temperature GKS. In high wavenumber region, TKE spectral from  high order GKS is closer to the experiment result, which outweighs results from second order GKS. High order accuracy is achieved in high order GKS,  which has advantage of simulating non-equilibrium multi-temperature flow when smooth equilibrium region appears. Besides, tiny difference resulting from the different bulk viscosity term between high order equilibrium GKS and high order GKS is observed in this TKE spectral. This different behavior is also verified by the PDF of the local Mach number $M_{loc}$ and the contours of $\Delta_T$ as the Figure \ref{dhit_pdf_dt}, as the maximum difference between translational temperature and rotational temperature on the $z = 0.5$ plane at dimensionless time $t^{\ast}=0.87$ is no more than $1.2 \%$.

\subsection{Low-density nozzle flow}
Low-thrust rocket engine has been used for the control of altitude and trajectory of satellites and spacecrafts. For this type of rocket engine, the fluid experiences continuum, transition flow regime, which provides a necessary test for the validity of current high order GKS method for near continuum flow regime.

Low density nozzle flow has been measured using the electron beam fluorescence technique by Rothe \cite{rothe1971electron}, and DSMC simulations have been performed by Chung et al. \cite{chung1995low}. The flow condition for the test case is stagnation temperature $T_0 = 300K$, stagnation pressure $P_0 = 474Pa$, wall temperature $T_w = 300K$. This is an axis-symmetric flow problem, only one quarter part of this nozzle has been computed with $340 \times 60 \times 60$ grid points used inside the nozzle. Empirical first-order slip boundary condition \cite{maxwell1879vii} is used in current high order GKS method for isothermal boundary condition.

\begin{figure}[!h]
	\begin{minipage}{0.46\linewidth}
		\centering
		\includegraphics[height=0.8\textwidth]{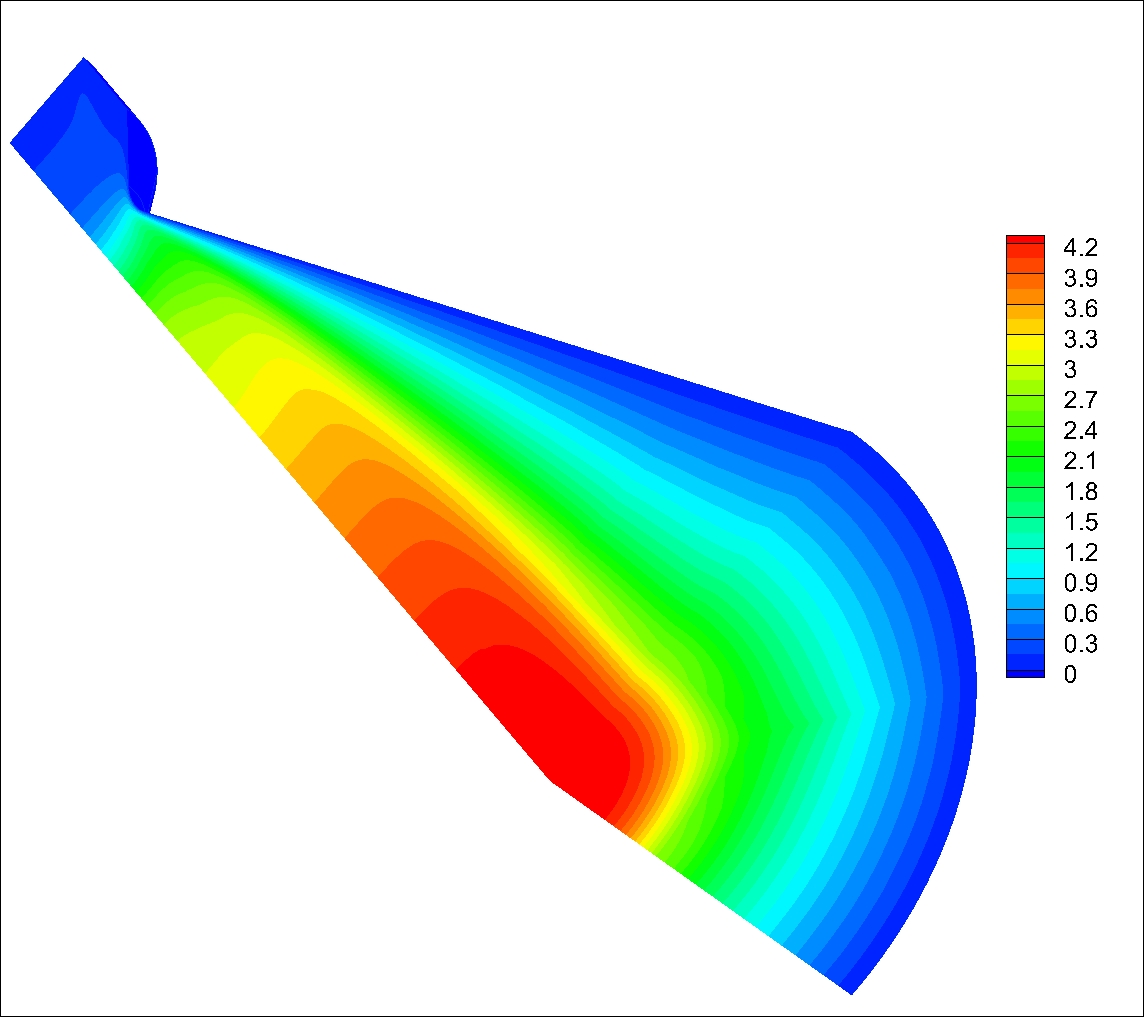}
	\end{minipage}
	\qquad
	\begin{minipage}{0.46\linewidth}
		\includegraphics[height=0.8\textwidth]{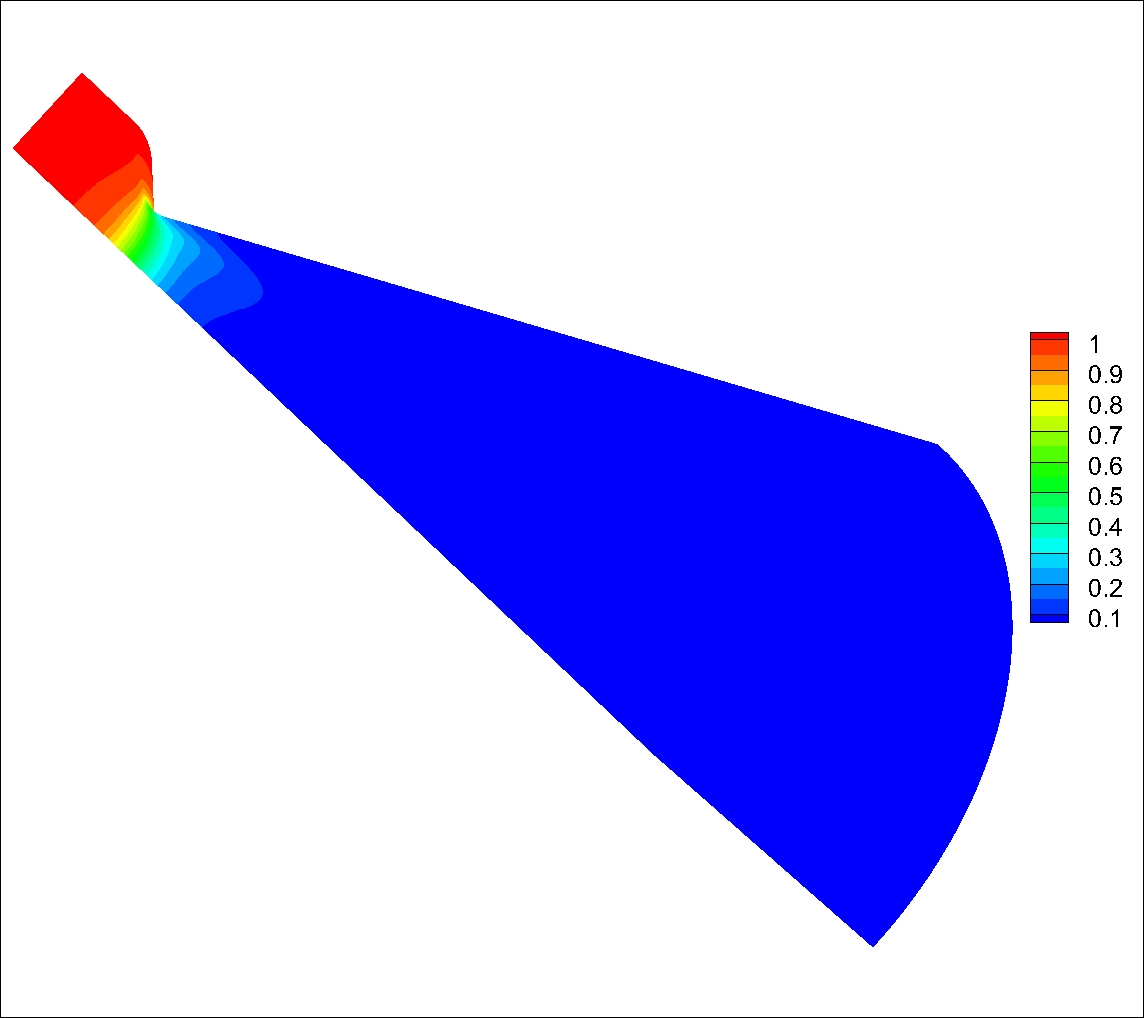}
	\end{minipage}
	\caption{Mach contour (left) and non-dimensional density contour (right) in the nozzle flow computations.}
	\label{nozzle_contour}
\end{figure}
\begin{figure}[!h]
	\begin{minipage}[c]{0.48\linewidth}
		\centering
		\includegraphics[height=0.9\textwidth]{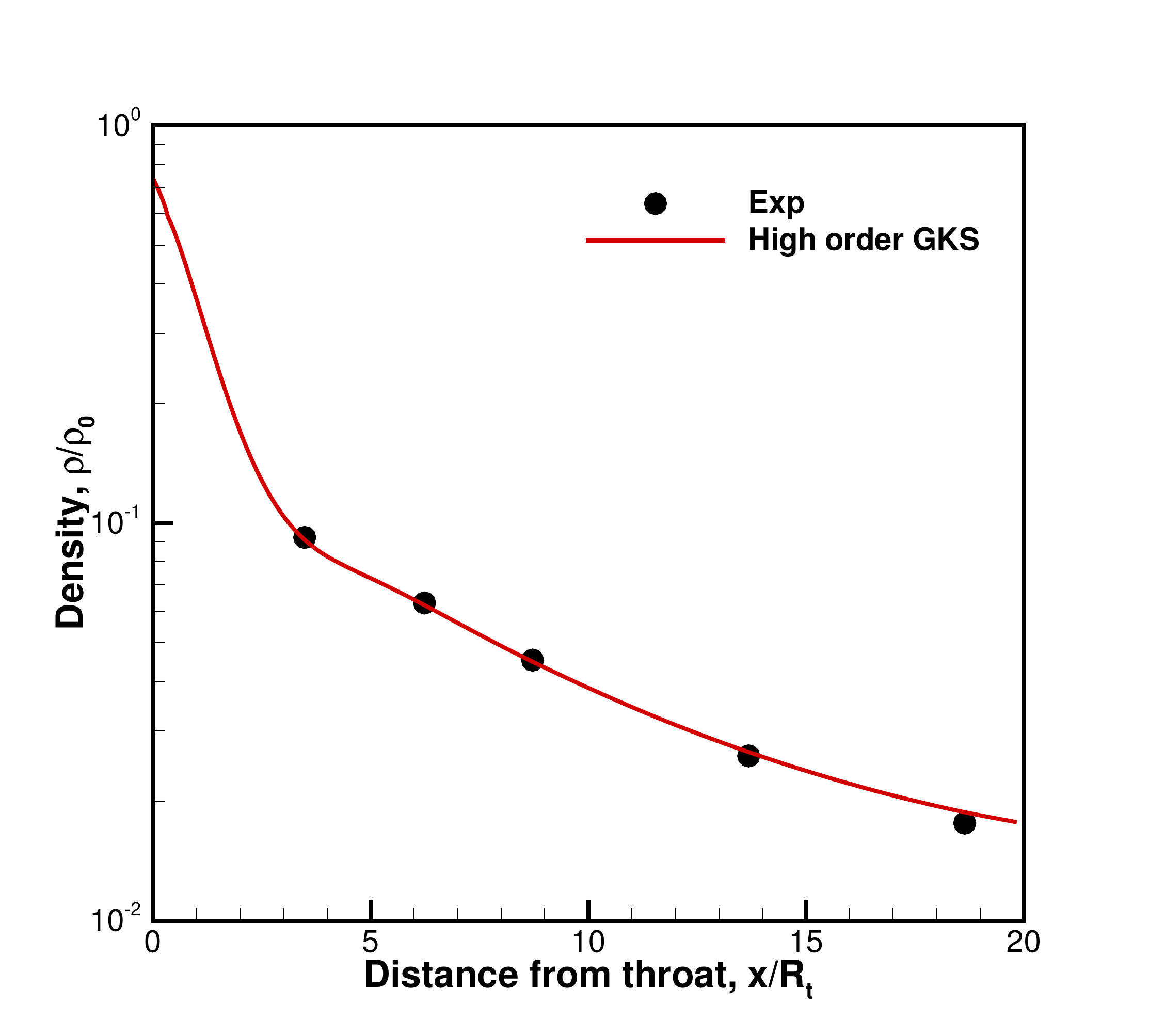}
	\end{minipage}
	\begin{minipage}{0.48\linewidth}
		\includegraphics[height=0.9\textwidth]{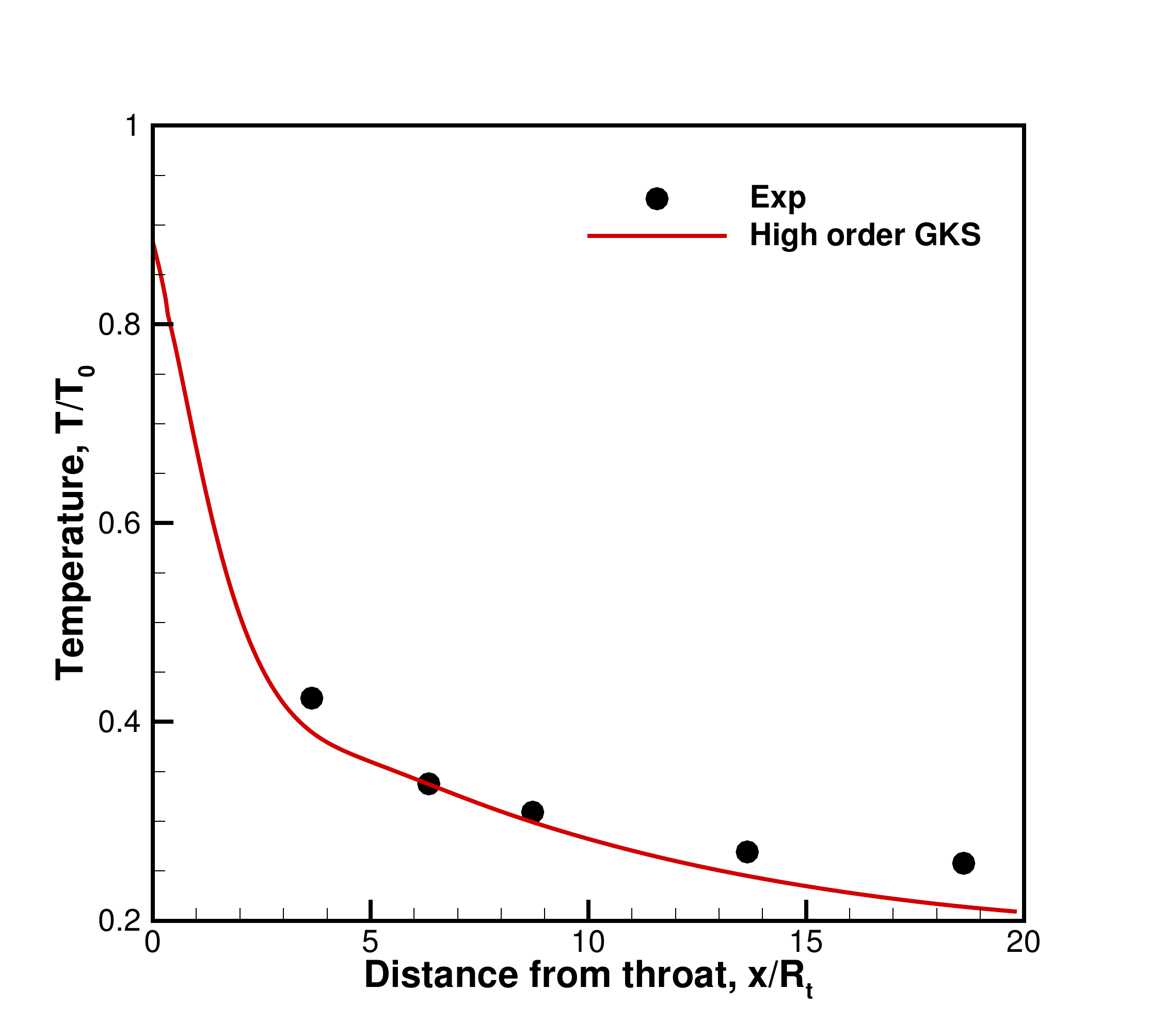}
	\end{minipage}
	\caption{Density and rotational temperature distributions along the central line of the nozzle, where $R_t$ is the throat radius. The measured rotational temperature is from experiment \cite{rothe1971electron}.}
	\label{nozzle_centerline}
\end{figure}

Figure \ref{nozzle_contour} shows the Mach contour and non-dimensional density contour inside this nozzle, where high ratio of density from inlet to outlet are observed. The experimental data of density and rotational temperature along the nozzle centerline are shown in Figure \ref{nozzle_centerline}. Current high order GKS method is validated in near continuum flow regime, as computation results provides a close match with the experimental measurement.

\subsection{Rarefied hypersonic flow over a flat plate}
\begin{figure}[!h]
	\begin{minipage}{0.46\linewidth}
		\centering
		\includegraphics[height=0.8\textwidth]{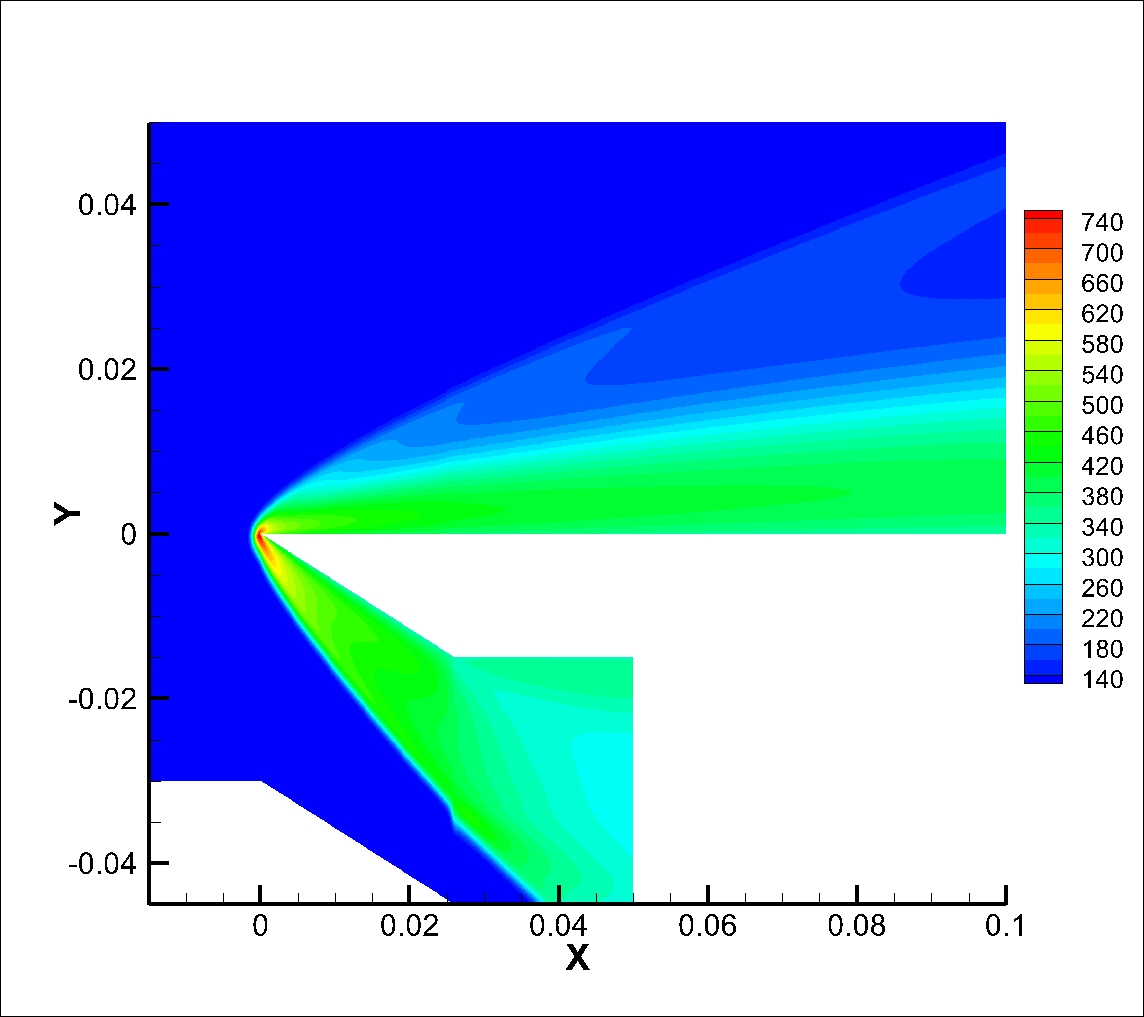}
	\end{minipage}
	\quad
	\begin{minipage}{0.46\linewidth}
		\includegraphics[height=0.8\textwidth]{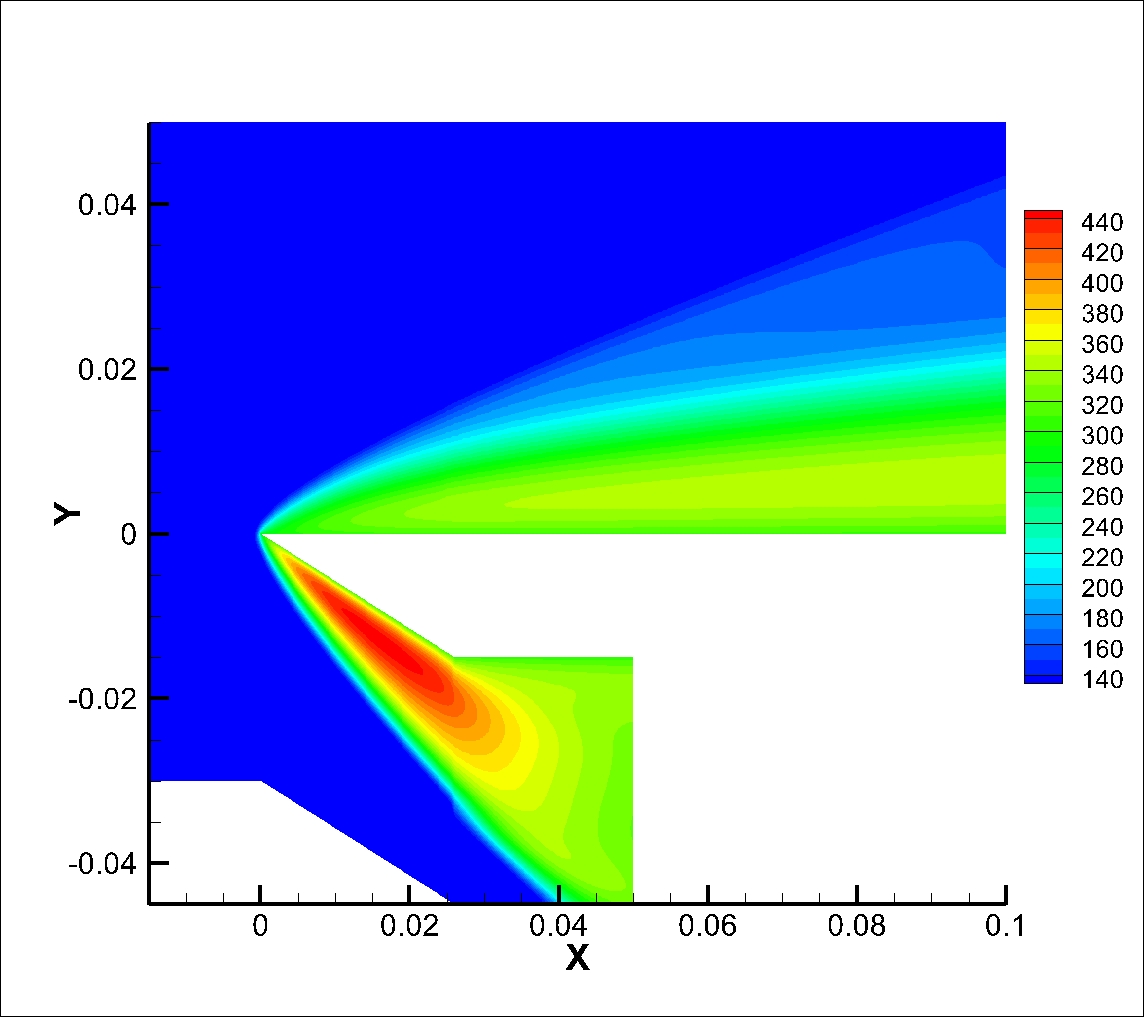}
	\end{minipage}
	\caption{Translational (left) and rotational (right) temperature contours in the hypersonic flow over a flat plate.}
	\label{plate_contour}
\end{figure}

Physical phenomena occurring around spacecraft in a hypersonic rarefied gas flow are studied in order to  understand these phenomena and to design a real size vehicle. Following the experiment conducted by Tsuboi et al \cite{tsuboi2005experimental}, simulation on the hypersonic rarefied gas flow over a flat plate is implemented. The case is the run 34, with the nozzle exit Mach number $Ma = 4.89$, stagnation temperature $T_0 = 670K$, stagnation pressure $P_0 = 983Pa$, nozzle exit temperature $T_e = 116K$, and flat plate surface temperature $T_w = 290K$ with first-order slip boundary condition used. The geometry is shown in Fig \ref{plate_contour}, where  $400\times 200$ and $300 \times 100$ grid points above and below the flat plate are used. In this case, the shock wave and boundary layer interaction near a sharp leading edge caused non-equilibrium between translational and rotational temperatures in the rarefied gas regime.

\begin{figure}[!h]
	\begin{minipage}{0.48\linewidth}
		\centering
		\includegraphics[height=0.9\textwidth]{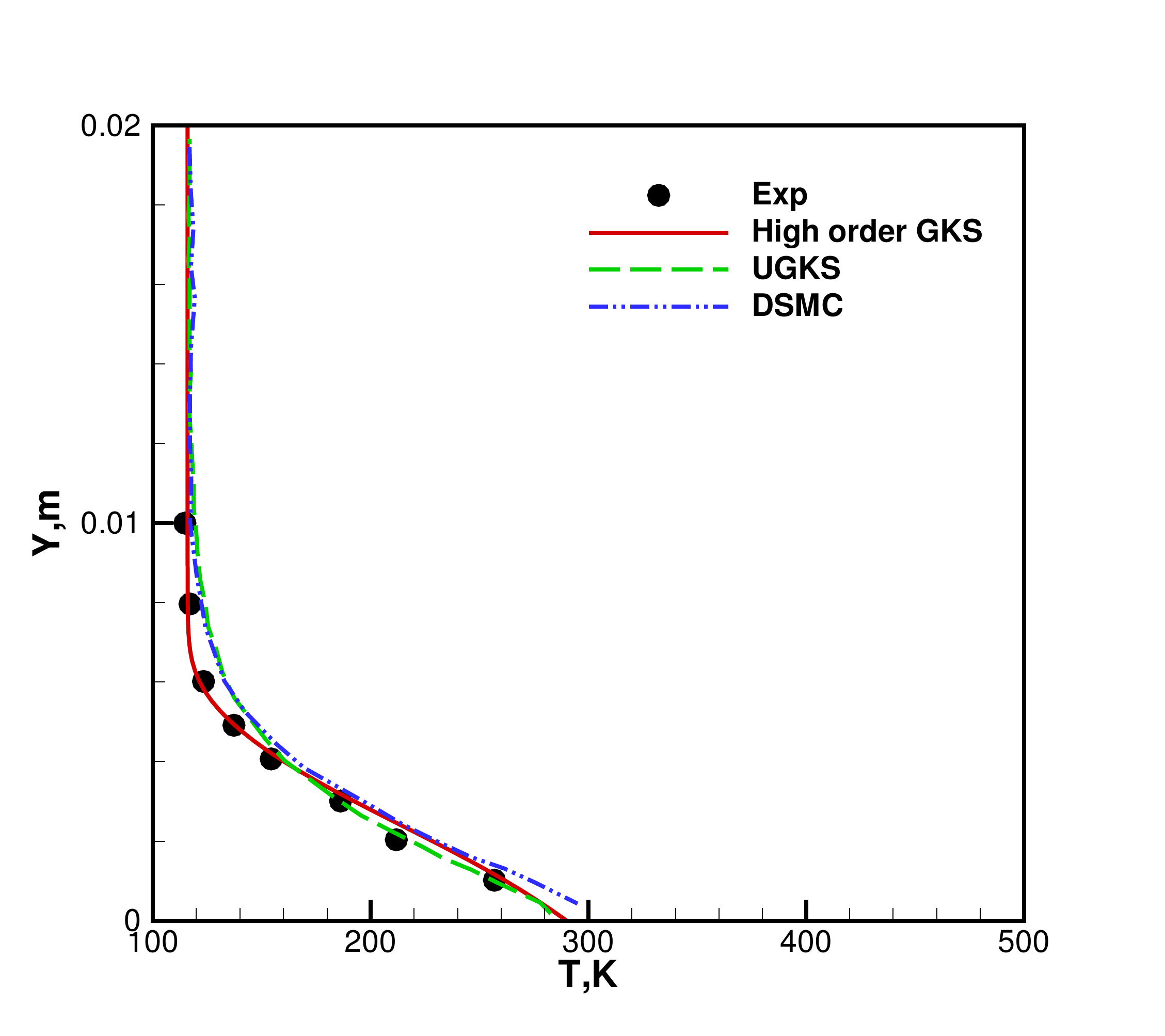}
	\end{minipage}
	\
	\begin{minipage}{0.48\linewidth}
		\includegraphics[height=0.9\textwidth]{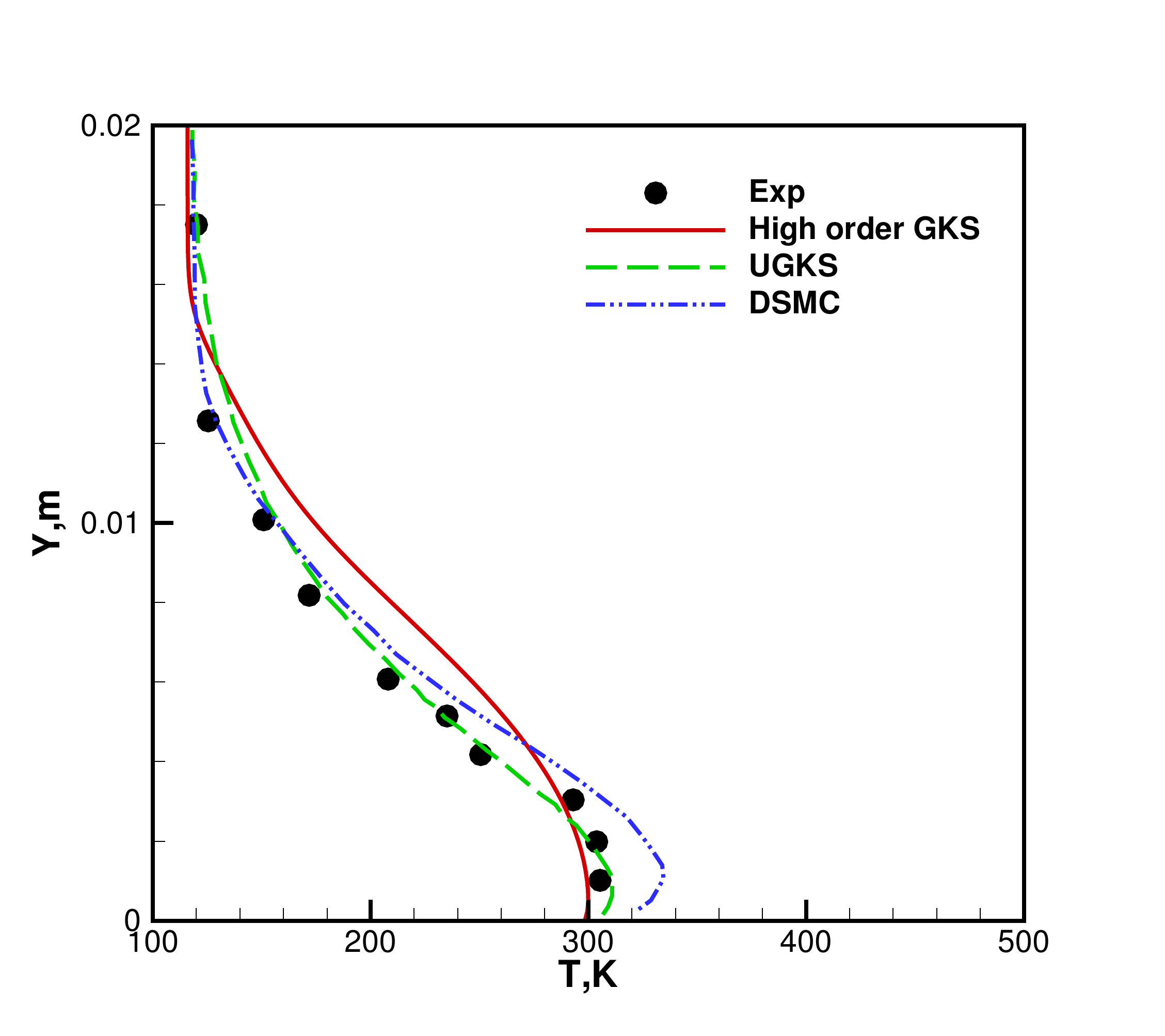}
	\end{minipage}
	\caption{Rotational temperature distributions in the vertical direction at $x = 5mm$ (left) and $x = 20mm$ (right). The measure rotational temperature \cite{tsuboi2005experimental}, current high order GKS solutions, UGKS solution \cite{liu2014unified}, and DSMC solution \cite{tsuboi2005experimental} are presented.}
	\label{plate_temperature_distribution}
\end{figure}

The temperature distributions in the vertical direction above the flat plate at the locations of $x = 5mm$ and $x = 20mm$ from the leading edge are shown in Figure \ref{plate_temperature_distribution}. As a comparison, the  UGKS results \cite{liu2014unified} and DSMC results \cite{tsuboi2005experimental} are also included. As shown in Figure \ref{plate_temperature_distribution}, current high order GKS result is comparable with DSMC result, while current high order GKS is more efficient than DSMC. While, UGKS results have a perfect match with the experiment measurement than current high order GKS method and DSMC solution, which shows its great advantage of multi-scale properties for the whole flow regime simulation. Here coarse grids in physical space is used in UGKS scheme, with $59 \times 39$ grid points above the plate and $44 \times 25$ below the plate. However, velocity space is discretized with $80 \times 60$ grid points in UGKS scheme, so current high order GKS method is still competitive in near continuum flow regime considering its higher efficiency than UGKS.

\subsection{Type IV shock-shock interaction}

Shock-shock interaction is the key issue in hypersonic flow. The presence of intense shock waves interaction strongly affects vehicle aerodynamic performance and leads to substantial localized aerodynamic heating. Shock-shock interaction was classified by Edney \cite{edney1968anomalous} into six patterns, depending on the impinging position and angle. In this paper, the type IV interaction is studied, which is the most severe case to form the hot spot on the surface of the cylinder due to the supersonic jet hitting on the wall. The flow patterns of the formation of a supersonic impinging jet, a series of shock waves, expansion waves, and shear layers in a local area of interaction, form a pretty challenging case for such a high-order GKS scheme.
\begin{figure}[h]
	\centering
	\includegraphics[height=0.35\textwidth]{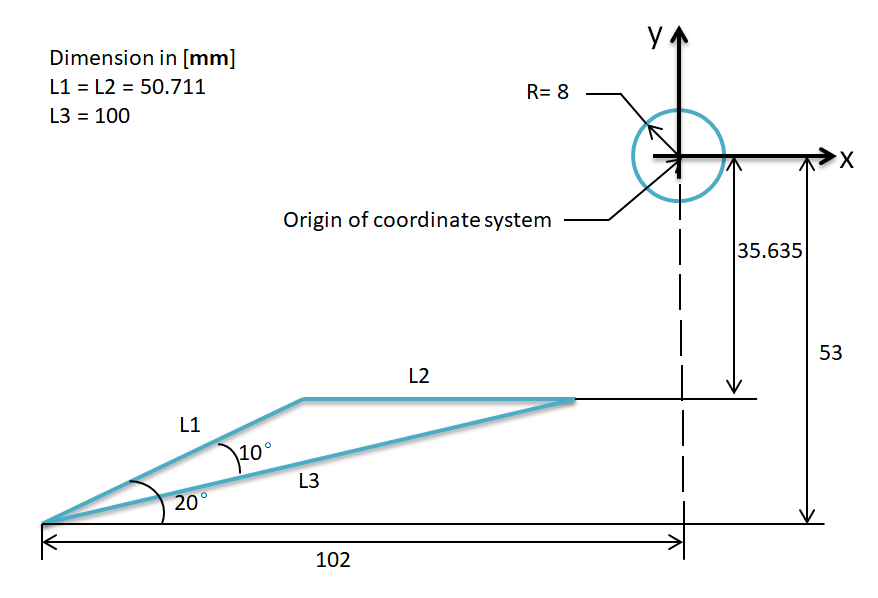}
	\centering
	\includegraphics[height=0.3\textwidth]{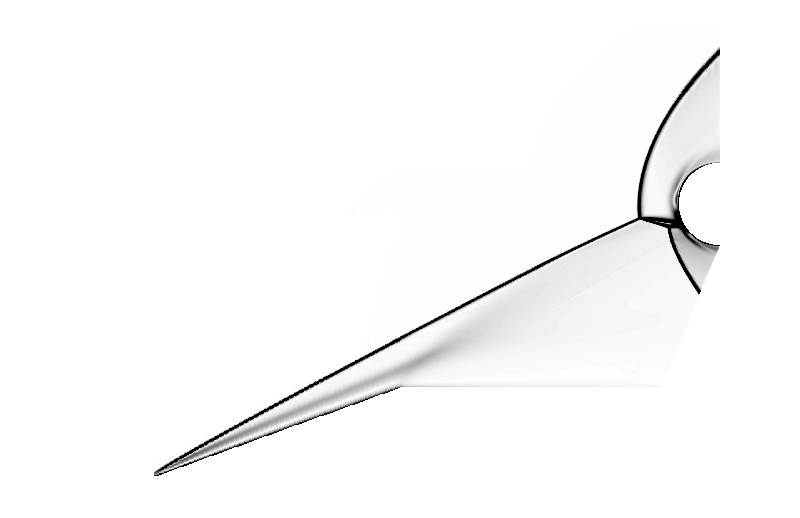}
	\caption{Configuration for ONERA experiment \cite{pot1998fundamental} (left) and Schlieren images by density gradient magnitude (right) from current high order GKS for type IV shock-shock interaction.}
	\label{edeyiv_configuration_schlieren}
\end{figure}

An experimental test has been conducted by Office national d'études et de recherches aérospatiales (ONERA) \cite{pot1998fundamental} to investigate shock-shock interactions, which provides free-stream air flow properties of $M_{\infty} = 10$, $T_{\infty} = 52.5K$, $T_w = 300K$, and $Re_{\infty}/m = 1.66 \times 10^5$. The leading edge of the shock generator is positioned at a distancel $L = 102mm$ upstream of the cylinder and $53mm$ below the axis of the cylinder, and the cylinder diameter is $16mm$. Our simulation is based on $250 \times 440$ grid points around the cylinder. Configuration for ONERA shock-shock interaction experiment and the Schlieren images by density gradient magnitude from current computational result are shown in Fig \ref{edeyiv_configuration_schlieren}. A steady state solution is obtained from the high order GKS scheme after a long time iteration with the iterative steps on the order of $10^5$ and the flow structure keeps the same form.
\begin{figure}[!h]
	\begin{minipage}{0.48\linewidth}
		\centering
		\includegraphics[height=0.9\textwidth]{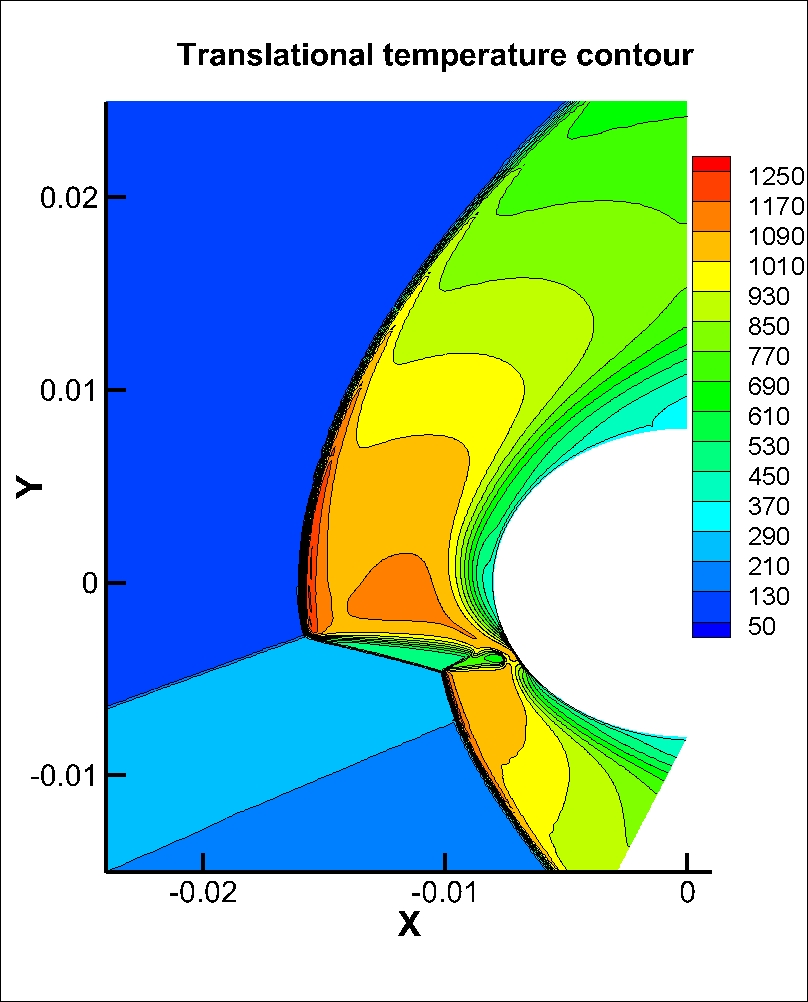}
	\end{minipage}
	\
	\begin{minipage}{0.48\linewidth}
		\includegraphics[height=0.9\textwidth]{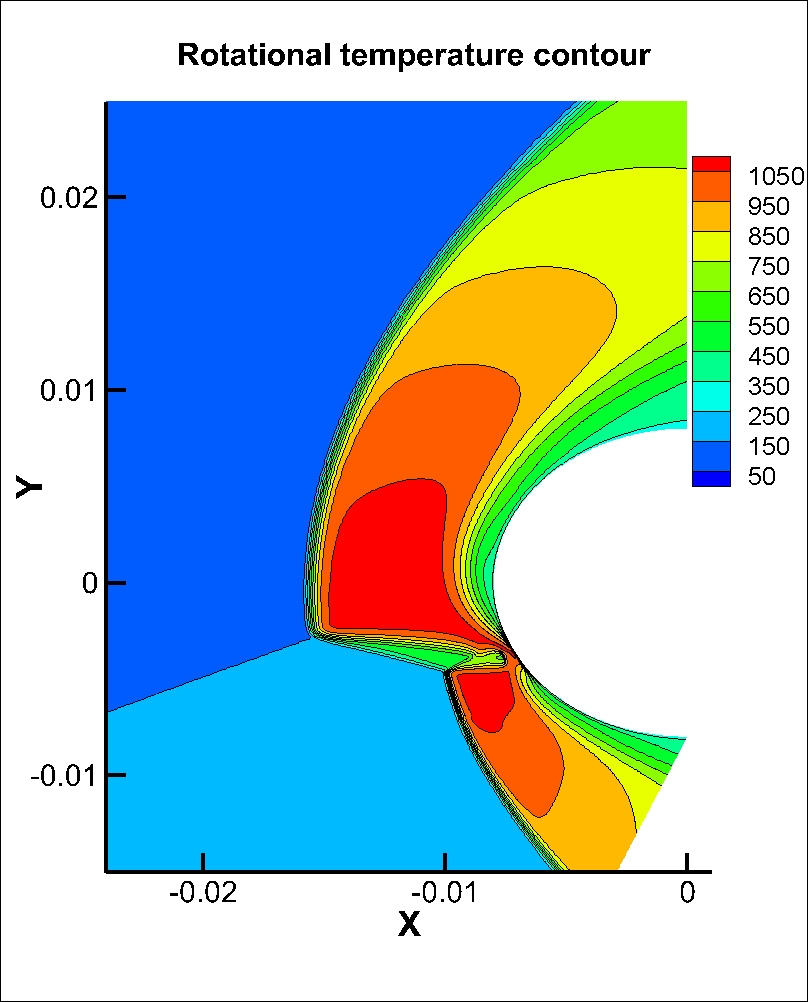}
	\end{minipage}
	\caption{Tranlational temperature contour (left) and rotational temperature contour (right) for type IV shock-shock interaction.}
	\label{edeyiv_temperature_contour}
\end{figure}
\begin{figure}[!h]
	\begin{minipage}{0.48\linewidth}
		\includegraphics[height=0.62\textwidth]{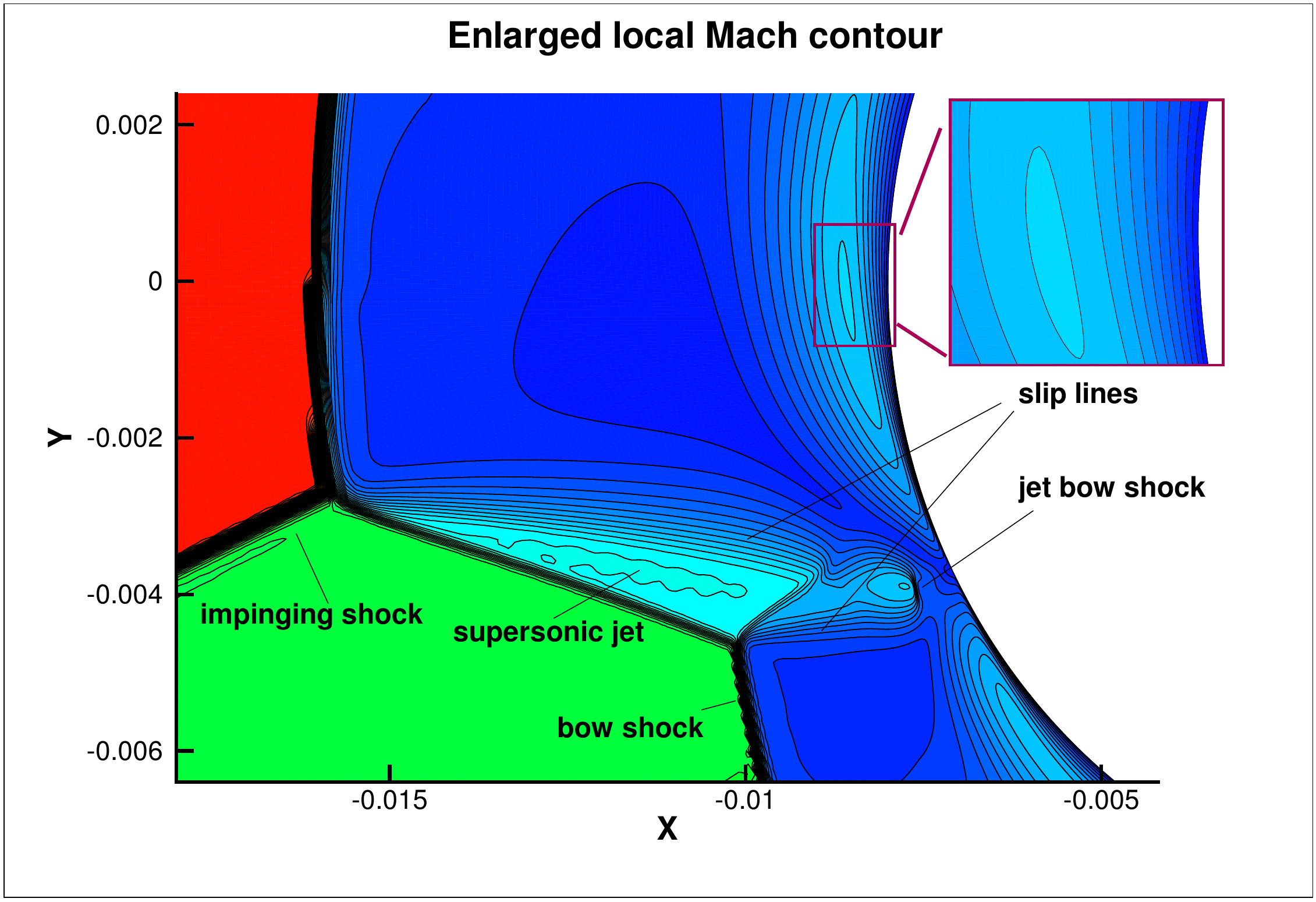}
	\end{minipage}
	\begin{minipage}{0.48\linewidth}
		\includegraphics[height=0.62\textwidth]{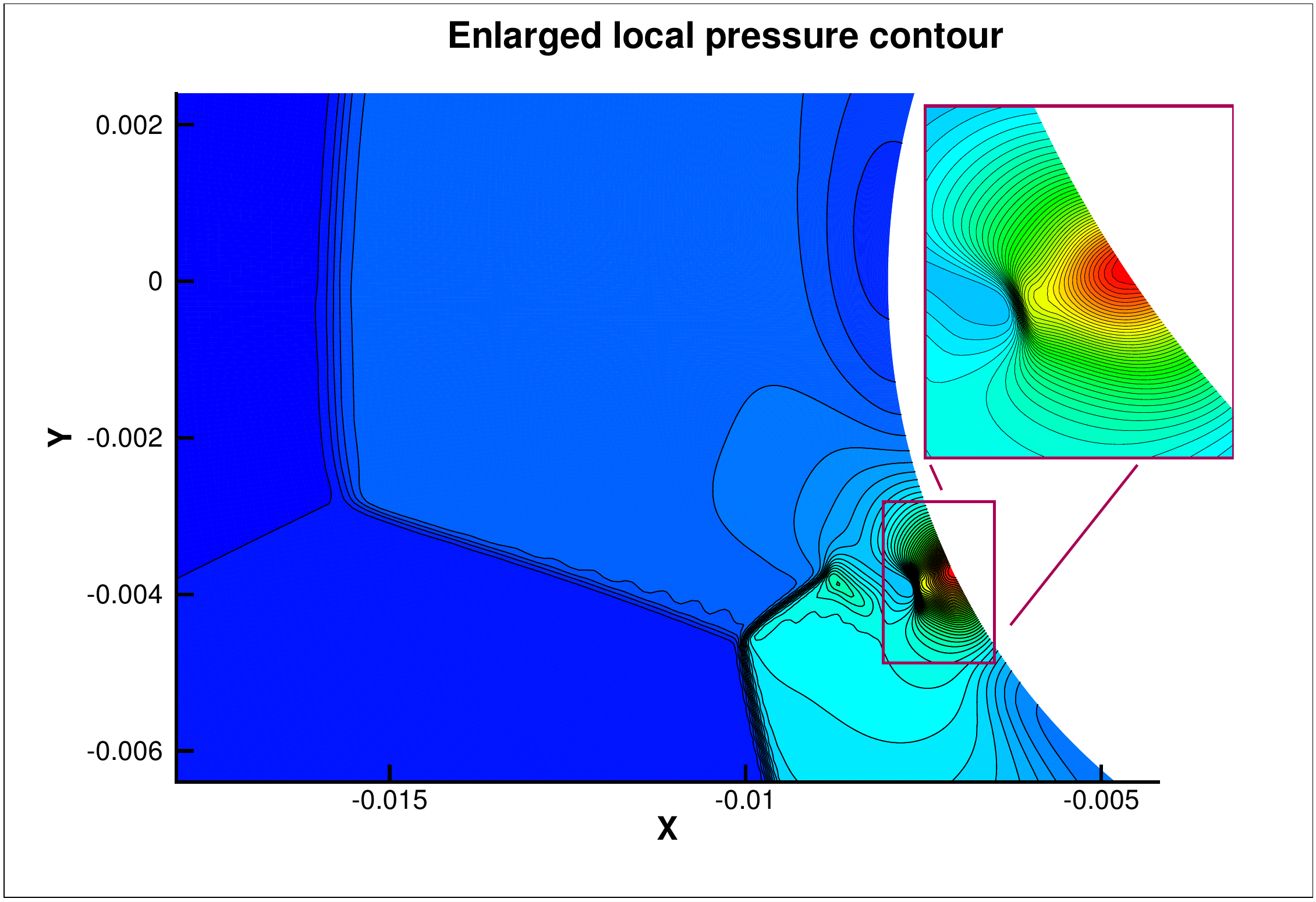}
	\end{minipage}
	\caption{Local Mach contour (left) and pressure (right) contour in the supersonic jet region.}
	\label{edeyiv_flowpattern}
\end{figure}

\begin{figure}[!h]
	\begin{minipage}{0.48\linewidth}
		\includegraphics[height=0.9\textwidth]{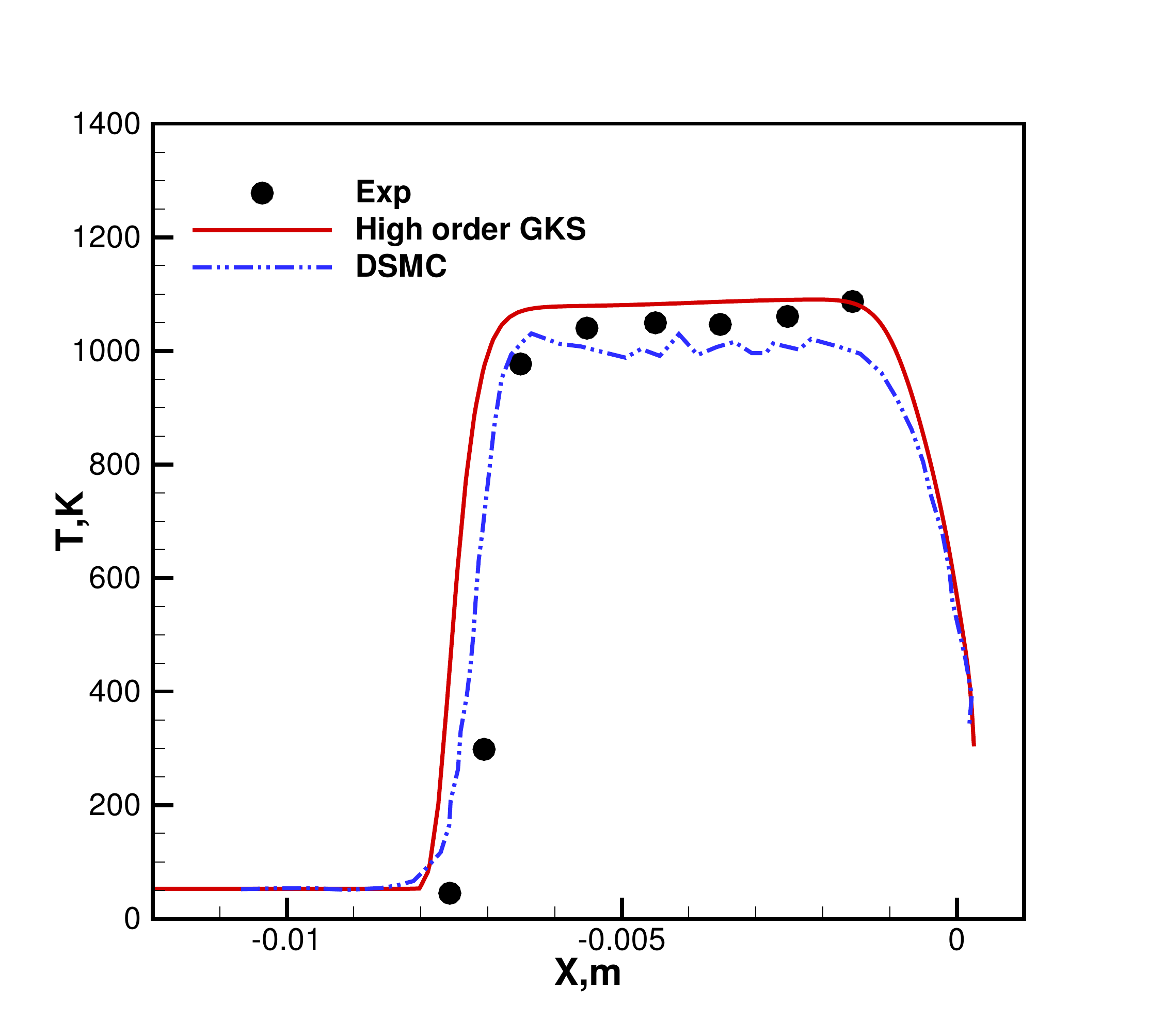}
	\end{minipage}
	\begin{minipage}{0.48\linewidth}
		\includegraphics[height=0.9\textwidth]{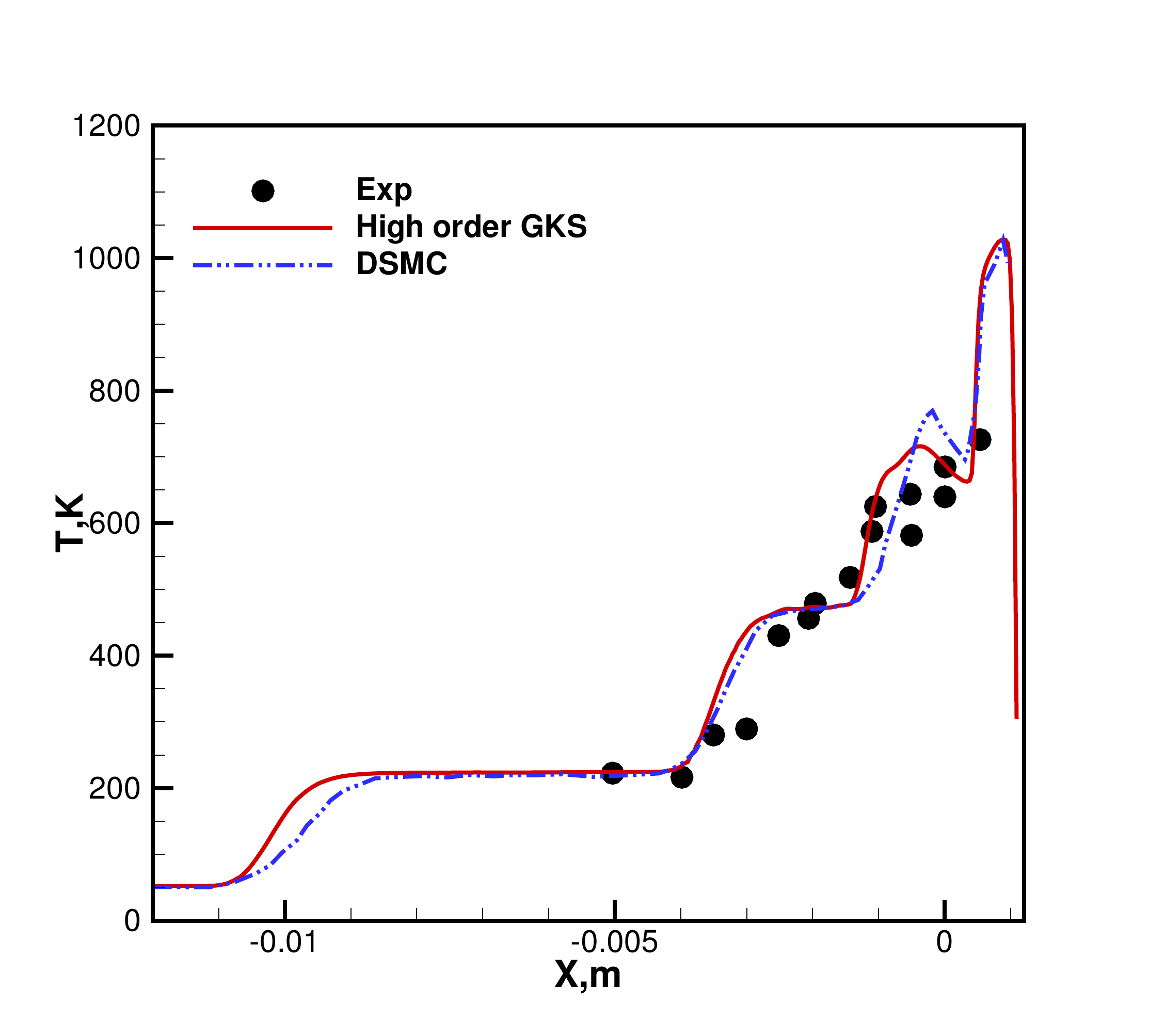}
	\end{minipage}
	\caption{Rotational temperature profile at $y = -2mm$ (left) and profile at $y = -4mm$ (right). The measured rotational temperature \cite{pot1998fundamental}, current high order GKS solutions, and DSMC solution \cite{moss1999dsmc} are presented.}
	\label{edeyiv_horizontal_profile}
\end{figure}

The translational temperature contour and rotational temperature contour around the cylinder are shown in Fig \ref{edeyiv_temperature_contour}. These contours confirm the existence of multiple temperature for this hypersonic flow. More specifically, the Mach number and pressure in the supersonic jet region are shown in Fig \ref{edeyiv_flowpattern}, which clearly shows the strong jet and hot spot around the cylinder surface. Figure \ref{edeyiv_horizontal_profile} presents two horizontal profiles of measured rotational temperature in experiment. One is located above the upper shock triple point at $y = -2mm$, and the other is the line at $y = -4mm$, which passes the transmitted shock and intersects with the surface one degree below the location of jet impingement. The high-order GKS results are close to DSMC solution \cite{moss1999dsmc} at $y = -2mm$, while oscillation appears in DSMC simulation. At $y = -4mm$, our computational results have a closer match with the experiment than DSMC solution, especially near $x = 0mm$ region.
\begin{figure}[!h]
	\begin{minipage}{0.48\linewidth}
		\includegraphics[height=0.9\textwidth]{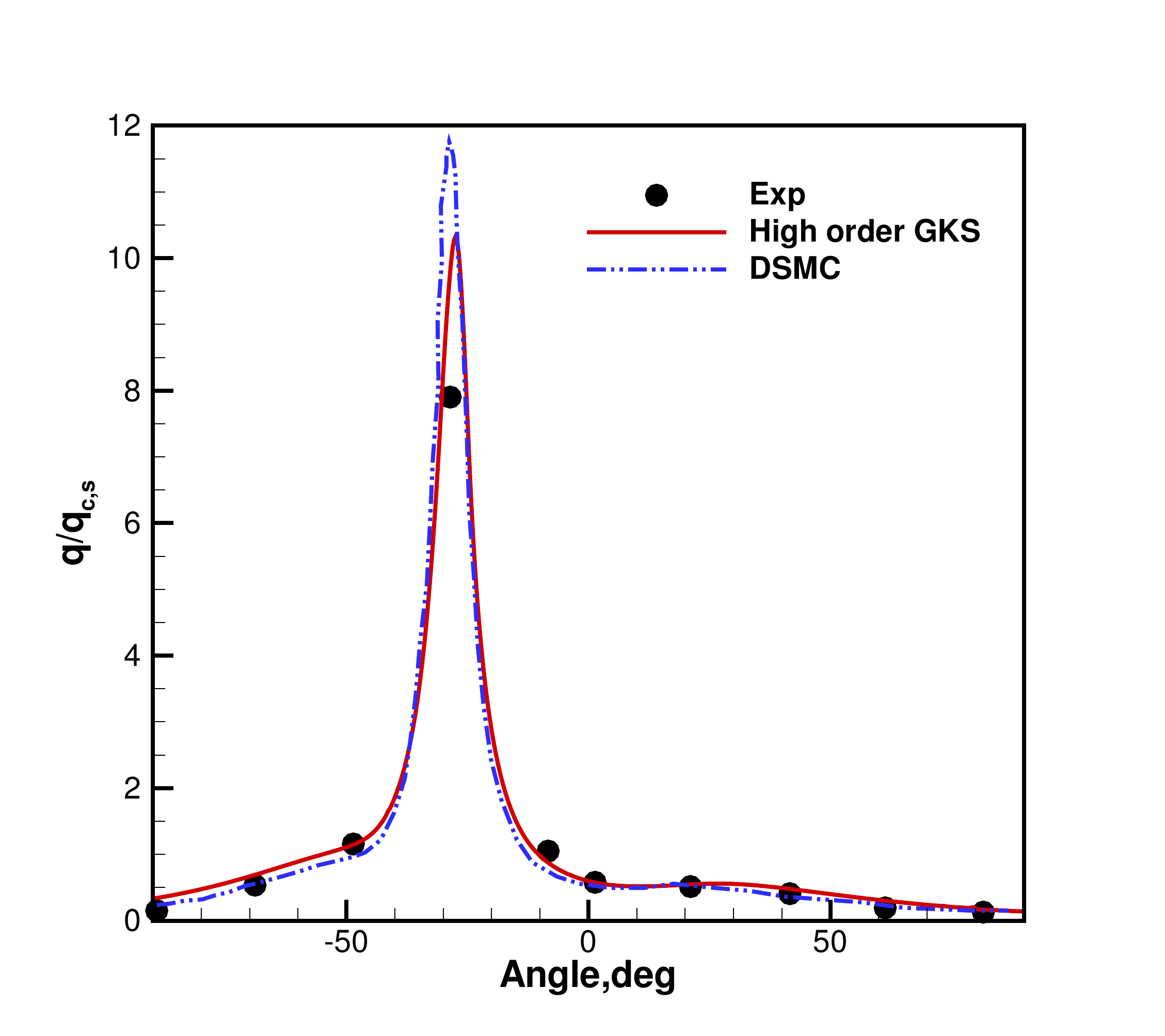}
	\end{minipage}
	\begin{minipage}{0.48\linewidth}
		\includegraphics[height=0.9\textwidth]{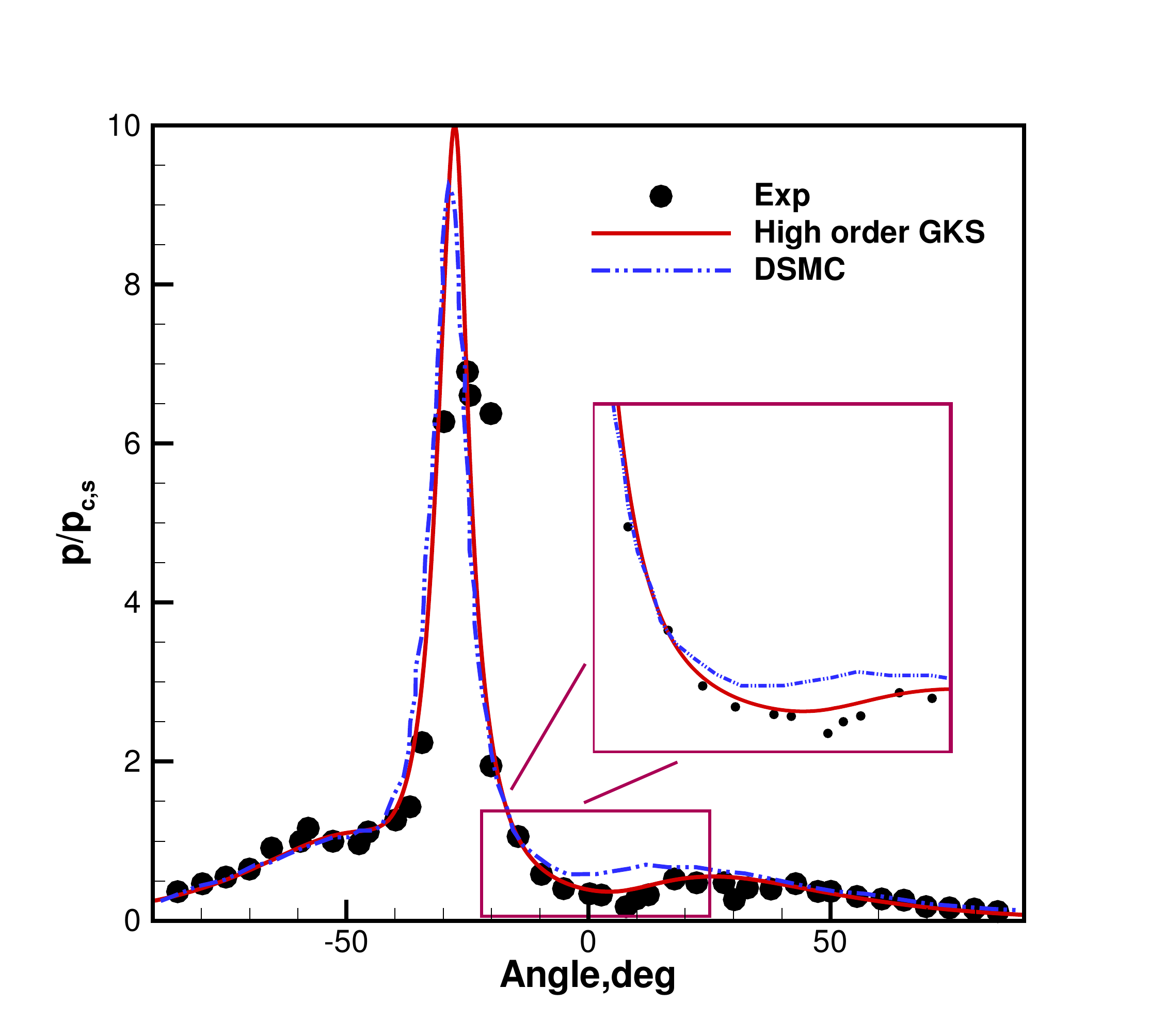}
	\end{minipage}
	\caption{Non-dimensional heating-rate distribution (left) and non-dimensional pressure distribution (right) along the cylindrical surface. The measured rotational temperature \cite{pot1998fundamental}, current high order GKS solutions, and DSMC solution \cite{moss1999dsmc} are presented.}
	\label{edeyiv_pressre_heat}
\end{figure}

The non-dimensional pressure and heat flux along the cylindrical surface from experimental measurements \cite{pot1998fundamental}, the high order GKS and DSMC computational results \cite{moss1999dsmc} are shown in Fig \ref{edeyiv_pressre_heat}, where $p_{c,s} = 760 Pa$ and $q_{c,s} = 5.7W/cm^2$ are the reference value for undisturbed flow about cylinder. The experimental heating data set are inadequate to define the peak value because of the limited spatial resolution, while the high order GKS and DSMC present the close peak position with different peak values. In terms of pressure distribution, the high order GKS outweighs DSMC results near $0^\circ$. Near $0^\circ$ region, a slightly low pressure region is found in Figure \ref{edeyiv_flowpattern}, which provides confidence on the high accuracy achieved by the high-order GKS scheme.

\section{Conclusion}
In this paper, a high-order three-dimensional multi-temperature GKS method is implemented under the two-stage fourth-order framework. Based on the extended BGK model, the three-dimensional macroscopic governing equations for diatomic gas are derived, which provide better insight into the behavior of the multi-temperature flow. Based on the developed multiple temperature kinetic model, a corresponding high-order GKS is constructed under the two-stage fourth-order framework and the source term discretization with fourth-order Simpson interpolation rule. For non-equilibrium multi-temperature flow computation, decaying homogeneous isotropic turbulence, nozzle flows, hypersonic rarefied flow over a plate, and type IV shock-shock interaction cases are tested. Comparisons among the numerical solutions from current high order GKS scheme, UGKS results, DSMC solutions, and experimental measurements show the high accuracy and quite robustness of current numerical method. Most importantly, the current finite volume gas-kinetic scheme updating the macroscopic flow variables explicitly, high efficiency is achieved in comparison with UGKS and DSMC methods, especially near the continuum flow region. \\

\section*{Appendix: Connection between BGK and Macroscopic Non-equilibrium Multi-temperature Equations in Three-dimensions}
Derivation of the Navier-Stokes and Euler equations from the BGK model can be found in the Appendix B in \cite{xu2015direct}. For macroscopic non-equilibrium multi-temperature equations in two-dimensions, it has been derived in \cite{xu2008multiple}. This appendix provides the details for the derivation to macroscopic non-equilibrium multi-temperature equations in three-dimensions. In this appendix, "Eq.(B.x)" represents the preliminary equation in Appendix B\cite{xu2015direct}, which will not be rewritten in current appendix. 

Continuity equation is given by 
\begin{align}
    \rho_{,t} + (\rho U_k)_{,k} = 0, \tag{A.1}
    \label{A1}
\end{align}
which can be used to simplify the momentum equations, the total energy equations, and the rotational energy energy equations. 

For momentum equations, the left side $\mathcal{L}_5$ in Eq.(B.2) can be grouped as
\begin{align*}
    \mathcal{L}_5 &= \frac{1}{2}U_n^2[{\rho_{,t} + (\rho U_k)_{,k}}] + \rho U_n U_{n,t} + \rho U_k U_n U_{n,k} + U_k p_{,k} \\
                  &+ \frac{K + 3}{2} [p_{,t} + U_k p_{,k}] + \frac{K + 5}{2} p U_{k,k} + {\frac{K}{2}(p_r - p)_t + \frac{K}{2}[(p_r - p) U_k]_{,k}}.
\label{A2}
\end{align*}
The first term is $\frac{1}{2} U_n^2 \mathcal{L}_1$ which is $\mathcal{O}(\epsilon^2)$, and next three are $U_n \mathcal{L}_n$, and are therefore $\mathcal{O}(\epsilon)$. Then $\mathcal{L}_5$ can be rewritten as 
\begin{align*}
    \mathcal{L}_5 &= \frac{K + 3}{2} [p_{,t} + U_k p_{,k}] + \frac{K + 5}{2} p U_{k,k} + U_n \mathcal{L}_n \notag \\
                  &+ {{\frac{K}{2}\{[(p_r)_t + (p_r U_k)_{,k}] - [p_{,t}  + (p U_k)_{,k}} ]\}}. \tag{A.2}
\end{align*}
Based on the Chapman-Enskog expansion up to zero order, rotational energy equation is obtained as
\begin{align*}
(\rho E_r)_t + (\rho E_r U_k)_{,k} = \frac{3 \rho}{2(K + 3) Z_r \tau}(\frac{1}{\lambda_t} - \frac{1}{\lambda_r}), \tag{A.3}
\end{align*}
which can be used to eliminate $(p_r)_t + (p_r U_k)_{,k}$. Based on $p_r = \rho E_r$, Eq.(A.2) can be rewritten as,
\begin{align*}
    -\frac{K + 3}{2} [p_{,t} + U_k p_{,k}] &= \frac{K + 5}{2} p U_{k,k} - {\color{black}{\frac{K}{2} p U_{k,k} }} \\
                                           &+ {\color{black}{\frac{K}{2}\{[\frac{3}{2(K + 3) Z_r \tau}(\frac{1}{\lambda_t} - \frac{1}{\lambda_r})] - [p_{,t} + U_k p_{,k}  } ]\}} + U_n \mathcal{L}_n + \mathcal{O}(\epsilon).
\end{align*}
Finally, we get
\begin{align*}
    p_{,t} + U_k p_{,k} = -\frac{5}{3} p U_{k,k} - \frac{K \rho}{2(K + 3) Z_r \tau}(\frac{1}{\lambda_t} - \frac{1}{\lambda_r}) + \mathcal{O}(\epsilon), \tag{A.4}
\end{align*}
which can be used to eliminate $p_{,t} + U_k p_{,k}$.

For the right sides of the momentum equations, we consider
\begin{align*}
    \mathcal{R}_j = (\hat{\tau} F_{jk})_{,k}.
\end{align*}
Using the fact that all odd moments in $w_k$ vanish, we get 
\begin{align*}
    F_{jk} &\equiv <u_j u_k>_{,t} + <u_j u_k u_l>_{,l} \\
           &= U_j[{\color{black}{(\rho U_k)_{,t} + [(\rho U_k U_l) + p \delta_{kl}]_{,l}}}] + \rho U_k U_{j,t} + (p \delta_{jk})_{,t} \\
           & + (\rho U_k U_l + p \delta_{kl}) U_{j,l} + (U_l p \delta_{jk} + U_k p \delta_{jl})_{,l}.
\end{align*}
The term in square brackets multiplying $U_j$ is $\mathcal{L}_k$, i.e. it is $\mathcal{O}(\epsilon)$, and can therefore be ignored. Then, after gathering terms with coefficients $U_k$ and $p$, we have
\begin{align*}
    F_{jk} = U_k[{\color{black}{\rho U_{j,t} + \rho U_l U_{j,l} + p_{,j}}}] + p[U_{k,j} + U_{j,k} + U_{l,l} \delta_{jk}] + \delta_{jk}[{\color{black}{p_{,t} + U_l p_{,l}}   }]. 
\end{align*}
The coefficient of $U_k$ is $\mathcal{L}_j$, according to Eq.(B.7), and can therefore be neglected. To eliminate $p_{,t}$ from the last term we use the Eq.(A.4) for $\mathcal{L}_5$. Finally, decompose the tensor $U_{k,j}$ into its dilation and shear parts in the usual way, which gives
\begin{align*}
    F_{jk} = [U_{k,j} + U_{j,k} - \frac{2}{3} U_{l,l} \delta_{jk}] - \frac{K \rho}{2(K + 3) Z_r}(\frac{1}{\lambda_t} - \frac{1}{\lambda_r}) \delta_{jk}. \tag{A.5}
\end{align*}

Analogy to derive the Navier-Stokes total energy equation, we write
\begin{align*}
    N_k \equiv <u_k \frac{u_n^3 + \xi_r^2}{2}>_{,t} + <u_k u_l \frac{u_n^3 + \xi_r^2}{2}>_{,l}.
\end{align*}
which can be written as 
\begin{align*}
    N_k = N_k^{(1)} + N_k^{(2)}.
\end{align*}
where 
\begin{align*}
    N_k^{(1)} =  [U_k \frac{u_n^3 + \xi_r^2}{2}]_{,t} + [U_k <u_l \frac{u_n^3 + \xi_r^2}{2}>]_{,l}.
\end{align*}
and
\begin{align*}
    N_k^{(2)} \equiv <w_k \frac{u_n^3 + \xi_r^2}{2}>_{,t} + <w_k u_l \frac{u_n^3 + \xi_r^2}{2}>_{,l} .
\end{align*}

For $N_k^{(1)}$, we have
\begin{align*}
    N_k^{(1)} &= U_k[{\color{black}{\frac{1}{2} <u_n^2 + \xi_r^2>_{, t} +  \frac{1}{2} <u_l (u_n^2 + \xi_r^2)>_{, l} }}] \\
    &+[\frac{1}{2} \rho U_n^2 + \frac{K + 3}{2} p] U_{k,t} + [U_l (\frac{1}{2}\rho U_n^2 + \frac{K + 5}{2} p)] U_{k,l} \\
    &+ { \color{black} {\frac{K}{2} (p_r - p) U_{k,t} + \frac{K}{2} U_l (p_r - p) U_{k,l}  } }.
\end{align*}
The coefficient of $U_k$ in the equation above is $\mathcal{L}_5$, and therefore can be dropped, and the remaining terms can be rewritten as
\begin{align*}
    N_k^{(1)} = [\frac{1}{2} \rho U_n^2 + \frac{K + 3}{2} p] [U_{k,t} + U_l  U_{k,l}] + p U_l U_{k,l} .
\end{align*}
According to equation Eq.(B.7) to eliminate $U_{k,t}$, we get
\begin{align*}
    N_k^{(1)} &= - [\frac{1}{2} U_n^2 + \frac{K + 3}{2} \frac{p}{\rho}] p_{,k} + p U_l U_{k,l} \\
              &+ { \color{black} {\frac{K}{2} (p_r - p) U_{k,t} + \frac{K}{2} U_l (p_r - p) U_{k,l}  } }. \tag{A.6}
\end{align*}

For $N_k^{(2)}$, remembering that moments odd in $w_k$ vanish, we have
\begin{align*}
N_k^{(2)} &= <U_n w_n w_k>_{,t} + <U_l U_n w_n w_k>_{,l} + \frac{1}{2} <U_n^2 w_k w_l>_{,l} + \frac{1}{2} <w_k w_l(w_n^2 + \xi_r^2)>_{,l} \\
          &= (p U_k)_{,t} + (p U_k U_l)_{,l} + \frac{1}{2}(U_n^2 p)_{,k} + \frac{K + 5}{2} (\frac{p^2}{\rho})_{,k} + \frac{K}{2} (\frac{p(p_r - p)}{\rho})_{,k} . 
\end{align*}

\newpage

This result can be written as
\begin{align*}
    N_k^{(2)} &= p [U_{k,t} + U_l U_{k,l} + U_k U_{l,l} + U_l U_{l,k}] \\
              &+ U_k(p_{,t} + U_l p_{,l}) + \frac{1}{2} U_n^2 p_{,k} + \frac{K + 5}{2} (\frac{p^2}{\rho})_{,k} \\
              &+{ \color{black} {\frac{K}{2} (\frac{p(p_r - p)}{\rho})_{,k}  } }.
\end{align*}
We want to eliminate the first order time derivative, so we rearrange above equality as
\begin{align*}
     N_k^{(2)} &= p [{\color{black}{U_{k,t}}} + U_l U_{k,l} + U_k U_{l,l} + U_l U_{l,k}] \\
               &+ U_k({\color{black}{p_{,t} + U_l p_{,l}}}) + \frac{1}{2} U_n^2 p_{,k} + \frac{K + 5}{2} (\frac{p^2}{\rho})_{,k} \\
               &+\frac{K}{2} (\frac{p(p_r - p)}{\rho})_{,k}.
\end{align*}
The $U_{k,t}$ can be eliminated by Eq.(B.7), and $p_{,t} + U_l p_{,l}$ can be eliminate by equation Eq.(A.4). Hence
\begin{align*}
    N_k^{(2)} &= p [U_k U_{l,l} - \frac{p_{,k}}{\rho} + U_l U_{l,k}] \\
              &+ U_k[{ \color{black}{ -\frac{5}{3}p U_{l,l} - \frac{K}{(K + 3) Z_r \tau}(p - p_r) } }] + \frac{1}{2} U_n^2 p_{,k}  \tag{A.7}\\
              &+ \frac{K + 5}{2} (\frac{p^2}{\rho})_{,k} + \frac{K}{2} (\frac{p(p_r - p)}{\rho})_{,k}.
\end{align*}

For $N_k$, sum up $N_k^{(1)}$ and $N_k^{(2)}$ together, obtaining\\
\begin{align*}
    N_k &=  p [U_l (U_{k,l} + U_{l,k}) - \frac{2}{3} U_k U_{l,l} ] - U_k \frac{K}{(K + 3) Z_r \tau}(p - p_r)  \\
        &+ { \color{black} \frac{K + 5}{2} p (\frac{p}{\rho})_{,k} + \frac{K}{2} (p_r - p) U_{k,t} + \frac{K}{2} U_l (p_r - p) U_{k,l} + \frac{K}{2} (\frac{p(p_r - p)}{\rho})_{,k}}.
\end{align*}
Eliminate $U_{k,t}$ by Eq.(B.7) again, leading to
\begin{align*}
    N_k &=  p [U_l (U_{k,l} + U_{l,k}) - \frac{2}{3} U_k U_{l,l} ] - U_k \frac{K}{(K + 3) Z_r \tau}(p - p_r)  \\
        &+  {\color{black}\frac{K}{2} p (\frac{p_r}{\rho})_{,k} + \frac{5}{2} p (\frac{p}{\rho})_{,k}}. \tag{A.8}
\end{align*}

For rotational energy equation, multiplying the continuity equation Eq.(A.1) by $\frac{K}{4 \lambda_r}$ and the subtracting the result from Eq.(A.3) gives, 
\begin{align*}
    \mathcal{L}_6 = \rho (\frac{K}{4 \lambda_r})_{t} + \rho U_k (\frac{K}{4 \lambda_r})_{k} - \frac{3 \rho K}{4(K + 3) Z_r \tau } (\frac{1}{\lambda_t} - \frac{1}{\lambda_r})+ \mathcal{O}(\epsilon^2). \tag{A.9}
\end{align*}
Unfolding $\mathcal{R}_6$, leads to
\begin{align*}
    \mathcal{R}_{6} &= \frac{\partial}{\partial x_k}\{\hat{\tau} [<\frac{1}{2} \xi_r^2 u_k>_{, t} + <\frac{1}{2} \xi_r^2 u_k u_l>_{, l}]\} \\
                    &= \hat{\tau} \{\frac{K}{4 \lambda_r} [{\color{black}{(\rho U_k)_{,t} + (\rho U_k U_l + p \delta_{k,l})_{,l} }}]  \\
                    &+ \rho U_k(\frac{K}{4 \lambda_r})_t + (\frac{K}{4 \lambda_r})_{,l} [\rho U_k U_l + p \delta_{kl}]\}_{,k}.
\end{align*}
The term in square brackets is $\mathcal{L}_k$, i.e. $\mathcal{O}(\epsilon)$, and can be dropped. Gathering terms with coefficients $U_k$ and $p$, and eliminating $\rho (\frac{K}{4 \lambda_r})_t + \rho U_l (\frac{K}{4 \lambda_r})_{,l}$ by Eq.(A.9), we have
\begin{align*}
    \mathcal{R}_{6} &= \hat{\tau} \{U_k [\rho (\frac{K}{4 \lambda_r})_t + \rho U_l (\frac{K}{4 \lambda_r})_{,l}] + (\frac{K}{4 \lambda_r})_{,l} p \delta_{kl}\}_{,k} \\
                    &=  \hat{\tau} \{U_k [\frac{3 \rho K}{4(K + 3) Z_r} (\frac{1}{\lambda_t} - \frac{1}{\lambda_r})] + (\frac{K}{4 \lambda_r})_{,l} p \delta_{kl}\}_{,k}. \tag{A.10}
\end{align*}
Above equations can be rewritten in the form of Eq.(\ref{macro_neq}). Hence, macroscopic non-equilibrium multi-temperature equations to three-dimensions have been derived.

\begin{acknowledgments}
	We would like to thank Xing Ji, providing the helpful discussion and suggestions. The authors would like to thank TianHe-II in Guangzhou for providing high performance computational resources. The current research is supported by HongKong research grant council (16207715, 16206617) and National Science Foundation of China (11772281, 91530319).
\end{acknowledgments}

\bibliography{caogybib}

\end{document}